\documentclass{aa}

\usepackage{graphicx}
\usepackage{txfonts}
\usepackage{hyperref} %[hidelinks] to remove the green boxs
\usepackage{float} % \begin{figure}[H] to force figures at a specific place
\usepackage{natbib}
\usepackage{multirow}
\usepackage{rotating}

\newcommand\Tstrut{\rule{0pt}{2.5ex}}         % = 'top' strut
\newcommand\Bstrut{\rule[-0.9ex]{0pt}{0pt}}   % = 'bottom' strut

\usepackage{xcolor}
\hypersetup
{
    colorlinks,
    linkcolor={red!50!black},
    citecolor={blue!70!black},
    urlcolor={blue!80!black}
}

\makeatletter
\renewcommand*\aa@pageof{, page \thepage{} of \pageref*{LastPage}}
\makeatother

\begin{document} 

   \title{Extremely metal-poor stars in the Fornax and Carina dwarf spheroidal galaxies\thanks{Based on UVES and X-Shooter observations collected at the ESO, program ID 0100.D--0820 and 094.D--0853.}}
   %\subtitle{}
   \author
   {    
        R. Lucchesi\inst{1,2}\fnmsep
        \and P. Jablonka\inst{2,4}
        \and Á. Skúladóttir\inst{1,5}
        \and C. Lardo \inst{6}
        \and L. Mashonkina\inst{7}
        \and F. Primas \inst{3}
        \and K. Venn \inst{8}
        \and V. Hill\inst{9}
        \and D. Minniti\inst{10,11,12}
   }

   \institute
   {
   Dipartimento di Fisica e Astronomia, Universit\'a degli Studi di Firenze, Via G. Sansone 1, I-50019 Sesto Fiorentino, Italy.
   \and Physics Institute, Laboratoire d'astrophysique, École Polytechnique Fédérale de Lausanne (EPFL), Observatoire, 1290 Versoix, Switzerland.
   \and European Southern Observatory, Karl-Schwarzschild-str. 2, 85748 Garching bei München, Germany.
   \and GEPI, Observatoire de Paris, Université PSL, CNRS, 5 Place Jules Janssen, 92190 Meudon, France.
   \and INAF/Osservatorio Astrofisico di Arcetri, Largo E. Fermi 5, I-50125 Firenze, Italy.
   \and Dipartimento di Fisica e Astronomia, Universit\`a degli Studi di Bologna, Via Gobetti 93/2, I-40129 Bologna, Italy.
   \and Institute of Astronomy of the Russian Academy of Sciences, Pyatnitskaya st. 48, 119017, Moscow, Russia.
   \and Department of Physics and Astronomy, University of Victoria, PO Box 3055, STN CSC, Victoria, BC V8W 3P6, Canada.
   \and Universit\'e C\^ote d’Azur, Observatoire de la C\^ote d’Azur, CNRS, Laboratoire Lagrange, Nice, France.
   \and Departamento de Ciencias F\'isicas, Facultad de Ciencias Exactas, Universidad Andr\'es Bello, Fern\'andez Concha 700, Las Condes, Santiago, Chile.
   \and Vatican Observatory, V00120 Vatican City State, Italy.
   \and Departamento de Fisica, Universidade Federal de Santa Catarina, Trinidade 88040-900, Florianopolis, Brazil.
    }

   \date{Received 28 September 2023 / Accepted 5 January 2024}

  \abstract 
  {We present our analysis of VLT/UVES and X-shooter observations of six very metal-poor stars, including four stars at [Fe/H]~$\approx$~$-3$  in the Fornax and Carina dwarf spheroidal (dSph) galaxies. To date, this metallicity range in these two galaxies has not yet been  investigated fully, or at all in some cases. The chemical abundances of 25 elements are presented, based on 1D and local thermodynamic
equilibrium (LTE) model atmospheres. We discuss the different elemental groups, and find that $\alpha$- and iron-peak elements in these two systems are generally in good agreement with the Milky Way halo at the same metallicity. 
  Our analysis reveals that none of the six stars we studied exhibits carbon enhancement, which is noteworthy given the prevalence of carbon-enhanced metal-poor stars without s-process enhancement (CEMP-no) in the Galaxy at similarly low metallicities.
  Our compilation of literature data shows that the fraction of CEMP-no stars in dSph galaxies is significantly lower than in the Milky Way, and than in ultra-faint dwarf galaxies. Furthermore, we report the discovery of the lowest metallicity, [Fe/H]~=~$-2.92$, r-process rich (r-I) star in a dSph galaxy. This star,  fnx\_06\_019, has $\rm[Eu/Fe]=+0.8$, and also shows enhancement of La, Nd, and Dy, $\rm[X/Fe]>+0.5$. Our new data in Carina and Fornax help  populate the extremely low metallicity range in dSph galaxies, and add to the evidence of a low fraction of CEMP-no stars in these systems.
  }

                
   \maketitle

\section{Introduction}
Our  aim is to  understand the characteristics of the first stars formed in the
universe from their imprints on low-mass stars in dwarf spheroidal 
(dSph) galaxies. The stellar abundance trends and dispersion of the most metal-poor
stars reveal the nature of the now disappeared first stellar generation  (e.g., mass,
numbers), and the level of homogeneity of the primitive interstellar medium
(ISM) (e.g., size and/or mass of star-forming regions), as well as the nature and energy of the first supernovae \citep[e.g.,][]{Koutsouridou2023}. The proximity of the Local Group dSph galaxies allows the  chemical abundances to be derived in individual stars at comparable quality to the Milky Way (MW). The confrontation of galaxies with very different evolutionary paths brings crucial information on the universality of the  star formation processes and chemical enrichment.

Carina, Sextans, Sculptor, and Fornax are four Local Group dSph galaxies that  have  
triggered strong observational efforts from the galactic
archaeology community. They provide  the first evidence for   star
formation histories and chemical evolution distinct from those of the Milky Way at
$\rm[Fe/H]>-2$ \citep[e.g.,][]{Tolstoy2009}.  These four dwarf galaxies form a
sequence of mass from the lowest to the highest mass limits of the classical
dSph galaxies. They followed very different evolutions: while the majority of stars in Sextans and Sculptor formed during the first 4\,Gyr \citep{deBoer2012, deBoer2014, Bettinelli2018, Bettinelli2019}, Carina is famous for its distinct star-forming episodes, which are separated by long periods of quiescence \citep{deBoer2014}.
With a significantly higher stellar mass, Fornax has an extended star formation history and a population dominated by intermediate-age stars \citep{deBoer2012}. Furthermore, in Fornax there are also six known globular clusters (GCs; \citealp{Hodge1961,Pace2021}).
The diversity of these four dSph galaxies allows us to probe the relation
between the very early stages of star formation and the subsequent evolutionary
paths.

While we now broadly understand their later stages of evolution, their earliest times are still poorly understood. Sculptor and Sextans are the
classical dSph galaxies with the largest number of metal-poor stars observed at sufficient spectral quality to derive accurate chemical abundances.
To date the most studied dwarf galaxy at low [Fe/H] is Sculptor, with 13 chemically analyzed metal-poor stars at [Fe/H]~$\le$~$-2.5$. \citep{Tafelmeyer2010,Frebel2010,Starkenburg2013,Jablonka2015,Simon2015,Skuladottir2021,Skuladottir2023b}. Only 4 extremely metal-poor (EMP) stars out of 14 known with $\rm [Fe/H]<-2.5$ have been studied in Sextans \citep{Shetrone2001,Aoki2009,Starkenburg2013,Lucchesi2020,Theler2020}. Fornax has only one known EMP star that has been chemically characterized, with total of two stars at [Fe/H]~$<-2.5$ \citep{Tafelmeyer2010, Lemasle2014}. Carina has nine stars with [Fe/H]~$<-2.5$ that have been chemically analyzed \citep{Koch2008,Venn2012, Susmitha2017, Norris2017,HansenT2023}.

\begin{figure}
\centering
\includegraphics[width=0.95\columnwidth]{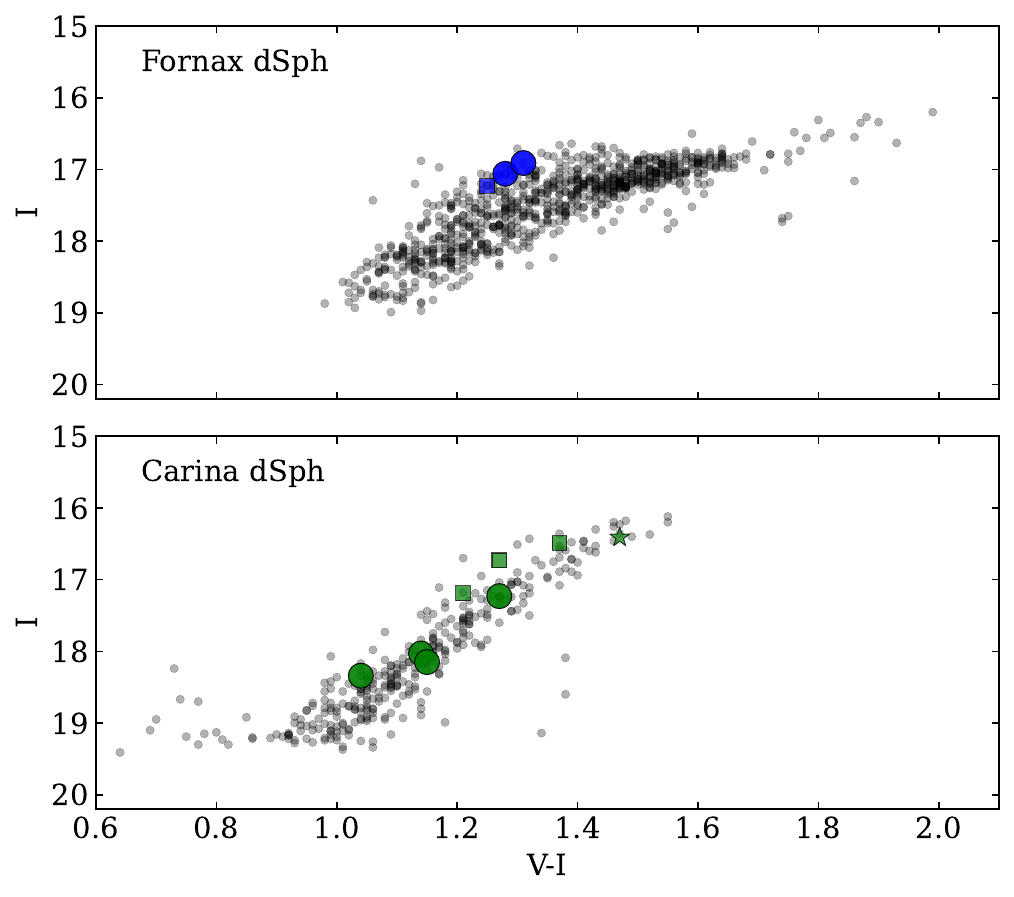}
\caption{Color--magnitude diagram ($I$, $V-I$) of the RGB of Fornax (top) and Carina (bottom). The $V, I$ magnitudes are from ESO 2.2m WFI \citep{Battaglia2006,Starkenburg2010}. The gray dots are probable Fornax and Carina members based on their radial velocities \citep{Starkenburg2010}. The colored points are stars with $\rm [Fe/H]\leq-2.5$ and more than five chemical abundances measured. The large blue and green  circles show respectively the Fornax and Carina stars analyzed here. Data shown from other works  are    \cite{Tafelmeyer2010} (small blue square);  \cite{Venn2012} (small green squares); and    \cite{Susmitha2017} (small green star).}
\label{Fig:CMD}
\end{figure}

\begin{figure}
\centering
\includegraphics[width=0.95\columnwidth]{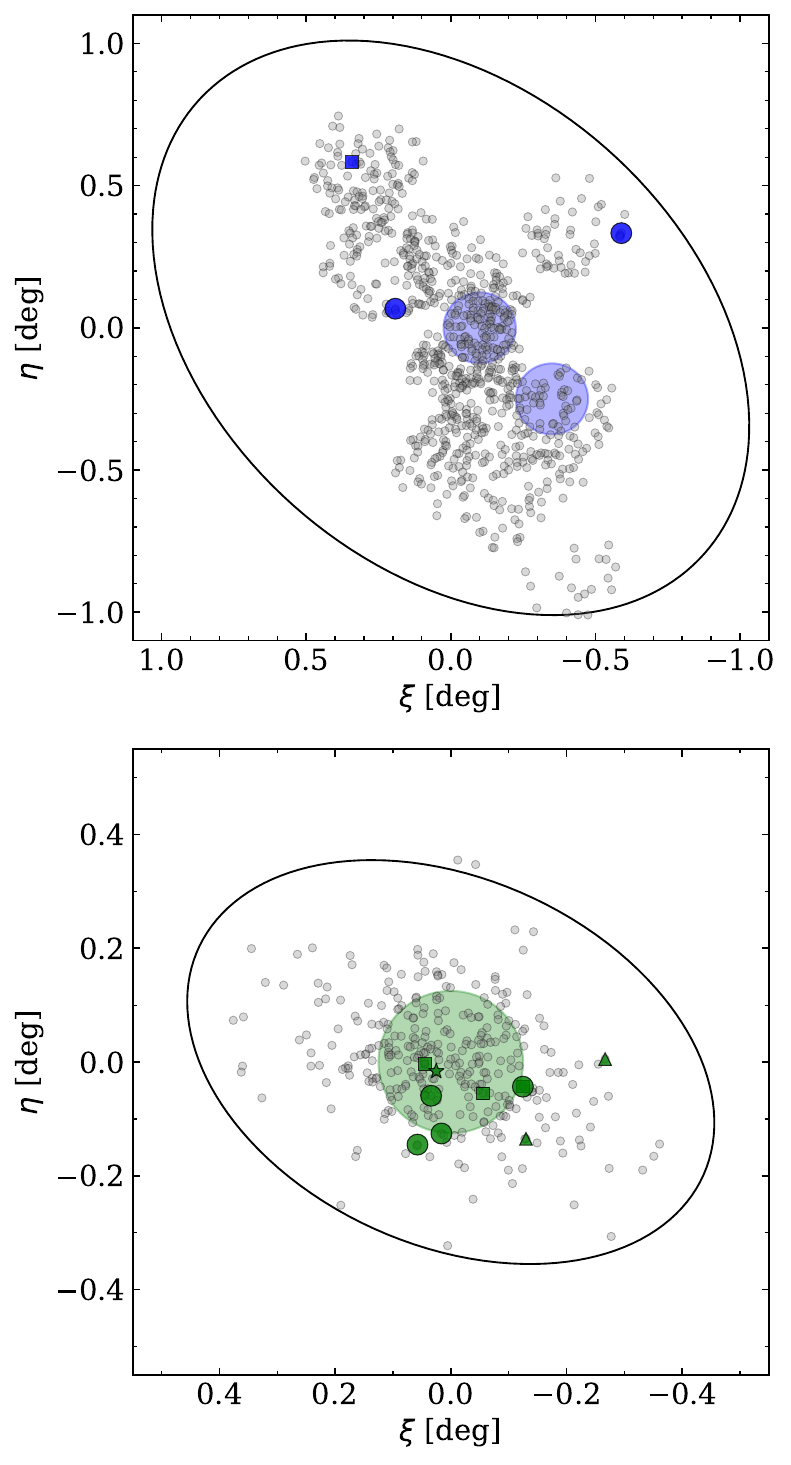}
\caption{Spatial distribution of stars. Top panel: Fornax stars, the large shaded blue circles correspond to the VLT/FLAMES observations of \cite{Letarte2010} (upper) and  of \cite{Lemasle2014} (lower). Bottom panel: Carina stars. The large shaded green circle corresponds to the VLT/FLAMES observations of \cite{Lemasle2012}, the green triangles are from \cite{Koch2008}; the other symbols are same as in Fig~\ref{Fig:CMD}. The ellipses indicate the tidal radius of Fornax and Carina, respectively.}
\label{Fig:map}
\end{figure}

\begin{table*}[htp]
\centering
\caption{\small Observation journal. The $\lambda$ range refers to the spectral ranges used in the analysis. The S/N is measured around the center of each wavelength range.}
\resizebox{0.8\linewidth}{!}{
\begin{tabular}{lcccccl}
\hline
\hline\Tstrut
ID & $\alpha (J2000) $ & $\delta (J2000) $  &  Setting & $\lambda$ range  & $<$S/N$>$ & V$_{rad, helio}$ $\pm~\sigma$  \\
   & $[$h:mn:s$]$      & $[ ^\circ$  :':"$]$ &          & \AA              & [/pix] & [km s$^{-1}$]  \\
\hline\Tstrut
 & & & \underline{UVES} & & &  \\
 & & & & & &   \\
fnx\_06\_019 & 02:37:00.91 & $-$34:10:43.10  & Dic1-CCD\#2       & 3800$-$4515 & 12 & 53.78~$\pm$~0.16 \\
            &             &                   & Dic1-CCD\#3(Blue) & 4790$-$5760 & 30 & 54.43~$\pm$~0.12 \\   
            &             &                   & Dic1-CCD\#3(Red)  & 5840$-$6805 & 45 & 53.98~$\pm$~0.16 \\
\hline\Tstrut
fnx0579x$-$1 & 02:40:47.79 & $-$34:26:46.50   & Dic1-CCD\#2       & 3800$-$4520 & 12 & 49.43~$\pm$~0.19 \\
             &             &                  & Dic1-CCD\#3(Blue) & 4790$-$5760 & 30 & 49.62~$\pm$~0.15 \\   
             &             &                  & Dic1-CCD\#3(Red)  & 5840$-$6805 & 43 & 49.17~$\pm$~0.18 \\
\hline
\hline\Tstrut
 & & & \underline{X-Shooter} & & &  \\
 & & & & & &  \\
car1\_t174  & 06:41:58.72 & $-$51:06:40.30 &      UBV          & 3040$-$5550 & 38 & 205.37~$\pm$~1.40 \\
            &             &                &      VIS          & 5550$-$6800 & 24 & 209.26~$\pm$~0.78 \\
\hline\Tstrut
car1\_t194  & 06:41:42.87 & $-$51:05:30.10 &      UBV          & 3040$-$5550 & 33 & 202.50~$\pm$~1.06 \\
            &             &                &      VIS          & 5550$-$6800 & 30 & 207.11~$\pm$~1.24 \\
\hline\Tstrut
car1\_t200  & 06:41:49.67 & $-$51:01:31.30 &      UBV          & 3040$-$5550 & 34 & 214.69~$\pm$~1.46 \\
            &             &                &      VIS          & 5550$-$6800 & 25 & 208.40~$\pm$~1.72 \\
\hline\Tstrut
LG04c\_0008 & 06:40:49.14 & $-$51:00:33.00 &      UBV           & 3040$-$5550 & 37 & 219.13~$\pm$~1.30 \\
            &             &                &      VIS           & 5550$-$6800 & 40 & 221.05~$\pm$~1.36 \\
\hline
\end{tabular}}
\label{Tab:journal}
\end{table*}

\begin{table}[htp]
\centering
\caption{Optical and near-IR photometry. $V$, $I$ from ESO 2.2m WFI. $J$, $H$, $K_s$ from ESO VISTA \citep{Battaglia2006,Starkenburg2010}.}
\resizebox{0.95\linewidth}{!}{
\begin{tabular}{cccccc}
\hline
\hline\Tstrut
    ID       & $V$    & $I$    & $J$    & $H$    & $K_s$ \\
\hline\Tstrut
fnx\_06\_019 & 18.336 & 17.062 & 16.005 & 15.455 & 15.310  \\
fnx0579x-1   & 18.220 & 16.910 & 16.090 & 15.532 & 15.393  \\
\hline\Tstrut
 car1\_t174   & 19.170 & 18.030 & 17.112 & 16.612 & 16.531  \\
 car1\_t194   & 19.300 & 18.150 & 17.203 & 16.689 & 16.584  \\
 car1\_t200   & 19.380 & 18.340 & 17.463 & 16.938 & 16.780  \\
LG04c\_0008   & 18.500 & 17.230 & 16.282 & 15.798 & 15.666  \\
\hline
\end{tabular}}
\label{Tab:photometry}
\end{table}

In  recent years it has become apparent that different dwarf galaxies show different abundance trends at low $\rm[Fe/H]<-2.5$. Around $\rm[Fe/H]\approx-3$, most dSph galaxies   (e.g., Fornax, Sextans, and Sculptor) \citep{Tafelmeyer2010,Theler2020,Lucchesi2020,Skuladottir2023b} present a plateau of enhanced [$\alpha$/Fe]$\sim$+0.4 with a low scatter, similar to the Milky Way and the small ultra-faint dwarf galaxies (UFDs). 
%This indicates a well-sampled initial mass function (IMF). 
However, there are some deviations from this, for example  in Sculptor, where individual unusual low-$\alpha$ stars have been detected at $\rm[Fe/H]\approx-2.4$ \citep[e.g.,][]{Jablonka2015}, and the [$\alpha$/Fe] plateau seems to break down toward the lowest $\rm[Fe/H]\approx-4$ \citep[e.g.,][]{Skuladottir2023b}. In addition, dSph  and UFD galaxies show significant differences in their Sr and Ba abundances. Typically, the values of  [Sr/Fe] and [Ba/Fe] in dSph galaxies increase toward higher [Fe/H], for example  in Sextans where $\rm[Sr/Fe]\approx0$ at $\rm[Fe/H]\gtrsim-3$, and $\rm[Ba/Fe]\approx0$ at $\rm[Fe/H]\gtrsim-2.6$ \citep{Lucchesi2020}. On the contrary, [Sr/Fe] and [Ba/Fe] typically remain subsolar at all [Fe/H] in UFDs \citep{Ji2019}.
The dSph galaxies show an increasing trend in [Sr/Ba] toward the lowest $\rm[Fe/H]\approx-4$. This is not seen in UFDs, and the Sextans and Ursa Minor dSph galaxies seem to be at the lower mass limit of where this trend is visible  \citep{Mashonkina2017,Reichert2020,Lucchesi2020}. It is likely that this relates to the earliest production sites of Sr and Ba; however, this is still under debated. 

The nucleosynthetic sites of the neutron-capture elements are still largely debated in the literature, in particular with regard to the rapid $r$-process, which occurs under high neutron flux \citep[e.g.,][]{Schatz2022}. Investigating stars in galaxies covering a wide mass range provides very crucial constraints, and supports a double origin of $r$-process elements \citep{SkuladottirSalvadori2020} by a rare type of massive stars \citep[e.g.,][]{Winteler2012}, as well as neutron star mergers \citep[e.g.,][]{Wanajo2014}.
In the Milky Way and and in dwarf galaxies individual stars have been found to exhibit enhancement of r-process products. These stars are defined as r-I stars when $\rm+0.3\le[Eu/Fe]<+1.0$, $\rm[Ba/Eu]<0$ 
 and as r-II stars when [Eu/Fe]$>$+1, [Ba/Eu]$<$0 \citep{Beers2005}. From the survey carried out by \citet{Barklem2005}, the expected frequency of r-II and r-I stars in the Milky Way halo is $\sim$3\% and $\sim$15\%, respectively. In Fornax, three r-II stars have been identified at high $\rm[Fe/H]>-1.3$ \citep{Reichert2021}, and several r-I stars have been reported at $\rm[Fe/H]>-1.5$  \citep{Letarte2010}. In the Ursa Minor dSph galaxy, r-II stars have also been identified at $\rm[Fe/H]>-2.5$ \citep{Shetrone2001,Aoki2007b}. In addition, some UFDs have been found to host very r-rich stars \citep{Ji2016a, HansenTT2017}. These metal-poor r-rich stars are extremely important since they likely show the imprints of a singular r-event. Understanding the distribution of such stars in different systems  therefore allows us to trace back the origin of the r-process enrichment. 

This work contributes to the study of the chemical patterns of the most
metal-deficient stars in dwarf galaxies, this time focusing on Fornax and
Carina.

%%%%%%%%%%%%%%%%%%%%%%%%%%%%%%%%%%%%%%%%%%%%%%%%%%%%%%%%%%%%%%%%%%%%%%%%%%%%%%%%%%%
\section{Observations and data reduction}\label{Sec:Observations}
%%%%%%%%%%%%%%%%%%%%%%%%%%%%%%%%%%%%%%%%%%%%%%%%%%%%%%%%%%%%%%%%%%%%%%%%%%%%%%%%%%%

\subsection{Target selection, observations, and data reduction}

The EMP candidates in this work are red giant branch (RGB) stars in the Fornax and Carina dSph galaxies (Figs.~\ref{Fig:CMD} and \ref{Fig:map}). They were 
selected   based on an estimation of their metallicity via the calcium~II triplet (CaT), [Fe/H]$_{{\rm CaT}} <$
--2.5. \citet{Starkenburg2010} delivered a CaT calibration down to $\rm[Fe/H]=-4$, which was applied to the   \citet{Battaglia2006} sample for Fornax and the  \citet{Koch2006} sample for Carina. The observational journal of the program  stars is in Table~\ref{Tab:journal}.

The two EMP candidates in Fornax, fnx\_06\_019 and fnx0579x$-$1, were
sufficiently bright to enable follow-up at high resolution with the UVES
spectrograph \citep{Dekker2000} mounted at the ESO VLT (program ID 0100.D-0820(A)). We
used dichroic\,1 with the CCD\#2 centered at 3900~\AA\ and the CCD\#3 centered
at 5800~\AA. Setting the slit width at 1.2\arcsec\ led to a nominal resolution of
R$\sim$34,000.  The full wavelength coverage is $\sim$3200--6800~\AA\ (see Table~\ref{Tab:journal} for details), and the
effective usable spectral information starts from $\sim$3800~\AA\ due to the lower signal-to-noise ratio (S/N) toward the blue part of the spectrum. 
Each star was observed for a total of five hours, split into six individual sub-exposures. The final S/N ranges from 12 (blue part) to 45 (red part).

The four EMP candidates in Carina (LG04c\_0008, Car1\_t200, Car1\_t174, and
Car1\_t194)  were observed (program ID 094.D--0853(B)) with  X-Shooter \citep{Vernet2011}. The
UVB slit was open to 0.8$\times$11 arcsec$^2$, while the VIS slit was open to
0.9$\times$11 arcsec$^2$, which led to a nominal resolution of R$\sim$6,200 and R$\sim$7,400, respectively. 
The total exposure time, in STARE mode, was 2.5 hours
for LG04c\_0008 and 3 hours for the other stars divided in 3 and 4 OBs of
$\sim$3000\,s, respectively. The usable wavelength range spans
3040--6800~\AA\ with S/N ranging from $\sim$24 to $\sim$40.

In all cases the reduced data, including bias subtraction, flat fielding,
wavelength calibration, spectral extraction, and order merging, were taken from
the ESO Science Archive Facility.
Table~\ref{Tab:journal} provides the coordinates of our targets, the
signal-to-noise ratios, and the radial velocities as measured in each
wavelength interval. Table~\ref{Tab:photometry} lists the optical and
near-infrared magnitudes of our sample. 
Figure~\ref{Fig:CMD} shows the color--magnitude diagram (CMD) of the targets, and  
Fig.~\ref{Fig:map} indicates the spatial location of our targets relative to 
the position of other spectroscopic studies in Carina and Fornax.

\subsection{Radial velocity measurements and normalization}\label{Sec:2.2}

The stellar heliocentric radial velocities (RVs) were measured with the {\tt
  IRAF}\footnote{Image Reduction and Analysis Facility; Astronomical Source Code
  Library ascl:9911.002} task {\it rvidlines} on each individual exposure. The
final RV is the average of these individual values weighted by their
uncertainties. This approach allows us to detect possible binary stars, at least
those whose RV variations can be detected within about one year.
We did not find any evidence for binarity. 
After they were corrected for RV shifts, the individual exposures were combined into a single spectrum using the {\tt IRAF} task {\it scombine} with a 2--3$\sigma$ clipping. As a final step, each spectrum was visually examined, and the few remaining cosmic rays were removed with the {\it splot} routine.

The mean RV of each Fornax star (Table~\ref{Tab:journal}) coincides with the RV of
Fornax, 54.1~$\pm$~0.5~$km s^{-1}$, within the velocity dispersion $\sigma =
13.7~\pm ~0.4~km s^{-1}$ measured by \cite{Battaglia2006}, and 55.46~$\pm$~0.63~$km s^{-1}$, within the velocity dispersion $\sigma =
11.62~\pm ~0.45~km s^{-1}$ measured by \cite{Hendricks2014b}. Similarly, the RVs of the Carina stars fall within the mean of the galaxy, 224.4~$\pm$~5.95~$km s^{-1}$ measured by \cite{Lemasle2012}, and 223.9~$km s^{-1}$ with $\sigma$=7.5~$km s^{-1}$ measured by \cite{Koch2006}. This confirms that our stars are galaxy members. The spectra were normalized using {\tt DAOSPEC} \citep{Stetson2008} for each of the wavelength ranges presented in Table~\ref{Tab:journal}. We used a 20 to 40 degree polynomial fit.

%%%%%%%%%%%%%%%%%%%%%%%%%%%%%%%%%%%%%%%%%%%%%%%%%%%%%%%%%%%%%%%%%%%%%%%%%%%%%%%%%%%
\section{Stellar model determination and chemical analysis}\label{Sec:analysis}
%%%%%%%%%%%%%%%%%%%%%%%%%%%%%%%%%%%%%%%%%%%%%%%%%%%%%%%%%%%%%%%%%%%%%%%%%%%%%%%%%%%

\subsection{Line list and model atmospheres}

Our line list combines those of \cite{Jablonka2015}, \citet{Tafelmeyer2010}, and
\citet{VanderSwaelmen2013}. Information on the spectral lines was taken from the
VALD database \citep{Piskunov1995,Ryabchikova1997,Kupka1999,Kupka2000}. The
corresponding central wavelengths and oscillator strengths are given in
Table~\ref{Tab:lines}.

We adopted the new MARCS 1D atmosphere models and selected the {\it Standard composition} class, that is, we included the classical $\alpha$-enhancement of +0.4~dex at low metallicity. They were downloaded from the MARCS web site \citep{Gustafsson2008}, and interpolated using Thomas Masseron's $interpol\_modeles$ code, which is available on the same web site.\footnote{\url{http://marcs.astro.uu.se}} Inside a cube of eight reference models, this code performs a linear interpolation on three given parameters: T$_{\mathrm{eff}}$, $\log$ g, and [Fe/H].

\subsection{Photometric temperature and gravity}\label{paramaters}

The atmospheric parameters were initially determined using photometric information, as reported in Table~\ref{Tab:photometry}.
The first approximated determination of the stellar effective temperature (T$_{\mathrm{eff}}$) was based on the V$-$I, V$-$J, V$-$H, and V$-$K color indices measured by \cite{Battaglia2006}, and the  J and Ks photometry was taken from the VISTA commissioning data, which were also calibrated onto the 2MASS photometric system.
We assumed $Av = 3.24 \cdot E_{B-V}$ \citep{Cardelli1989} and $E_{B-V}$~=~0.03 for Fornax \citep{Letarte2010}, and  $E_{B-V}$~=~0.061 for Carina \citep{deBoer2014} for the reddening correction. The adopted photometric effective temperatures ($T_{\mathrm{eff}}$) are listed in Table~\ref{Tab:parameters}. They correspond to the simple average of the four color temperatures derived from $V-I$, $V-J$, $V-H$, and $V-K$ with the calibration of \cite{Ramirez2005}.

\begin{table*}[ht]
\centering
\caption{CaT metallicity estimates, and  photometric and final spectroscopic parameters.}
\resizebox{0.9\textwidth}{!}{
\begin{tabular}{cccccccc|cccc}
\hline
\hline\Tstrut
              & \multicolumn{5}{c}{Photometric Parameters} &  &  &  \multicolumn{4}{c}{Final Parameters}\\
              & \multicolumn{5}{c}{T$_{\mathrm{eff}}$ [K]} &  &  & T$_{\mathrm{eff}}$ & $\log$(g) & v$_\mathrm{t}$ & [Fe/H] \\
  ID          & $V-I$ & $V-J$& $V-H$ & $V-K_s$ & mean $\pm$ $\sigma$ &$\log$(g) &    [Fe/H]$_{CaT}$     & [K]   & [cgs] &  [km s$^{-1}$] & \\
\hline\Tstrut
fnx\_06\_019  &  4379 & 4257 & 4248 & 4362  & 4311 $\pm$ 68 & 0.70 & $-2.54$   & 4280  & 0.68 & 1.80 & $-$2.92 \\
fnx0579x-1    &  4338 & 4422 & 4386 & 4427  & 4393 $\pm$ 41 & 0.71 & $-2.55$   & 4255  & 0.62 & 1.70 & $-$2.73 \\
\hline\Tstrut
 car1\_t174   &  4711 & 4678 & 4592 & 4646  & 4657 $\pm$ 51 & 1.42 & $-3.41$   & 4650  & 1.42 & 1.90 & $-$3.01 \\
 car1\_t194   &  4619 & 4533 & 4519 & 4544  & 4554 $\pm$ 45 & 1.42 & $-2.68$   & 4550  & 1.43 & 1.71 & $-$2.58 \\
 car1\_t200   &  4862 & 4813 & 4690 & 4663  & 4757 $\pm$ 96 & 1.56 & $-3.26$   & 4750  & 1.56 & 1.69 & $-$2.95 \\
LG04c\_0008   &  4518 & 4504 & 4466 & 4507  & 4499 $\pm$ 23 & 1.07 & $-3.29$   & 4520  & 1.08 & 1.78 & $-$3.05 \\
\hline
\end{tabular}}
\label{Tab:parameters}
\end{table*}

Because   very few \ion{Fe}{ii} lines can be detected in our X-Shooter spectra, the determination of surface gravities from the ionization balance of \ion{Fe}{I} versus \ion{Fe}{II} was not possible. Non-local thermodynamic equilibrium (NLTE) effects also play a role at extremely low metallicity, and impact the abundances of \ion{Fe}{I}, with $\Delta$(\ion{Fe}{II}--\ion{Fe}{I}) up to +0.20~dex at [Fe/H]~=~$-3$ \citep{Mashonkina2017a}, and thus surface gravities were determined from their relation with $T_{\mathrm{eff}}$,
\begin{equation}
 \log g_{\star} = \log g_{\odot} + \log \frac{M_{\star}}{M_{\odot}} + 4 \times \log \frac{T_{{\mathrm{eff}} \star}}{T_{{\mathrm{eff}} \odot}} + 0.4 \times \left( M_{{\rm bol} \star}- M_{{\rm bol} \odot} \right)
 \label{Eq:logg}
,\end{equation}
assuming $\log g_{\odot}$~=~4.44, $T_{{\mathrm{eff}}\odot}$~=~5790 K, and M$_{\rm bol \odot}$~=~4.75 for the Sun. We adopted a stellar mass of 0.8 M$_{\odot}$ and calculated the bolometric corrections using the \cite{Alonso1999} calibration, with a distance of
d=138\,kpc \citep{Battaglia2006} for Fornax and d=106\,kpc \citep{deBoer2014} for Carina.

\subsection{Final stellar parameters and abundance determinations}

We determined the stellar chemical abundances via the measurement of the
equivalent widths (EWs) or spectral synthesis of atomic transition lines
when necessary (see below).
Lines present in the spectra and in our line list were detected and their EWs measured with {\tt DAOSPEC} \citep{Stetson2008}.
This code performs a Gaussian fit of each individual line
and measures its corresponding EW. Although {\tt DAOSPEC} fits saturated
Gaussians to strong lines, it cannot fit the wider Lorentz-like wings of the
profile of very strong lines, in particular beyond 120~m\AA\ at very high
resolution \citep{Kirby2012}. For some of the strongest lines in our
spectra, we therefore derived the abundances by spectral synthesis.
All other detected lines by {\tt DAOSPEC} were also inspected visually and the measured EWs were rejected when too uncertain (for example in the case of upper limits the abundances were determined through spectral synthesis), or the lines were totally rejected when the continuum level was too uncertain.

The measured EWs are provided in Table~\ref{Tab:lines} and \ref{Tab:lines2}. The values in parentheses
indicate that the corresponding abundances were derived by spectral synthesis.
The abundance derivation from EWs and the spectral synthesis calculation were
performed with the {\tt Turbospectrum} code \citep{Alvarez1998,Plez2012}, which
assumes local thermodynamic equilibrium (LTE), but treats continuum scattering
in the source function. We used a plane-parallel transfer for the line
computation; this is consistent with our previous work on EMP stars
\citep{Tafelmeyer2010,Jablonka2015,Lucchesi2020}.

In order to derive the final T$_{\mathrm{eff}}$ and the microturbulence
velocities ($v_\mathrm{t}$), we checked or required no trend between the
abundances derived from \ion{Fe}{i} and excitation potential ($ \chi_{exc} $) or
the {\em \textup{predicted}}\footnote{According to \citet{Magain1984}, the use of observed EWs produce an
  increase in $v_\mathrm{t}$ of 0.1$-$0.2 $km s^{-1}$, which would be reflected
  in a decrease in the measured [Fe/H] values of a few hundredths of a dex in a
  systematic way. A variation like this does not change the results    significantly.} EWs. In order to minimize the
NLTE effect on the measured abundances we excluded from the
analysis \ion{Fe}{i} lines with $\chi_{exc}$~<~1.4~eV. Furthermore, very strong lines (EW~>~120m\AA) with strong wings that cannot be well fit are also excluded.

Starting from the initial photometric parameters of Table~\ref{Tab:parameters}, we adjusted
$T_{\mathrm{eff}}$ and $v_\mathrm{t}$ by minimizing the slopes of the diagnostic plots, within its 2$\sigma$ uncertainty.
We did not force ionization equilibrium between \ion{Fe}{I} and \ion{Fe}{II}, taking into account that there will likely be NLTE effects at these low metallicities \citep{Amarsi2016,Mashonkina2017a, Ezzeddine2017}. For each iteration the corresponding values of $\log g$ were computed from its relation with $T_{\mathrm{eff}}$ (Eq.~\ref{Eq:logg}), assuming the updated values of $T_{\mathrm{eff}}$, and adjusting the model metallicity to the mean iron abundance derived in the previous iteration.
The final values of T$_{\mathrm{eff}}$ are less than 30\,K away from the initial photometric estimates;  the only exception is fnx0579x$-$1, which is 138\,K cooler than the mean photometric temperature.

We derived the chemical abundances of blended lines, strong lines (EW~>~100~m\AA), or upper limits (EW~<~25~m\AA) by spectral synthesis. 
These abundances were obtained using our
own code (as in \citealt{Lucchesi2020,Lucchesi2022}), which performs a $\chi^2$-minimization between the observed spectral
features and a grid of synthetic spectra calculated on the fly with {\tt Turbospectrum}. 
A line of a chemical element $X$ is synthesized in a
wavelength range of $\sim$10 to $\sim$50~\AA. It is optimized by varying its abundance in
steps of 0.1~dex, from [$X$/Fe]~=~$-2.0$~dex to [$X$/Fe]~=~$+2.0$~dex. In the
same way, the resolution of the synthetic spectra is optimized when
needed. Starting from the nominal instrumental resolution, synthetic spectra
can be convolved with a wide range of Gaussian widths for each abundance step. A
second optimization, with abundance steps of 0.01~dex, is then performed in a
smaller range around the minimum $\chi^2$ in order to refine the
results. Similarly, the elements with a significant hyperfine structure (HFS; e.g., Sc, Mn, Co, and Ba) are determined by running {\tt Turbospectrum} in its
spectral synthesis mode in order to properly take into account blends and the
HFS components in the abundance derivation, as in \cite{North2012} and 
\cite{Prochaska2000} for Sc and Mn, and from the Kurucz web
site\footnote{\url{http://kurucz.harvard.edu/linelists.html}} for Co and Ba.

The final abundances are listed in Table \ref{Tab:abundances}. The solar abundances are taken from \cite{Asplund2009}.

\subsection{Error budget}\label{Sec:errors}

The uncertainties on the abundances were derived considering the uncertainties
on the atmospheric parameters and on the EWs, in a   procedure similar to that used in our other works \citep[e.g.,][]{Tafelmeyer2010,Jablonka2015,Hill2019,Lucchesi2020}.

{\em Uncertainties due to the atmospheric parameters.}
  To estimate the sensitivity of the derived abundances to the adopted atmospheric parameters, we repeated the abundance analysis and varied only one stellar atmospheric parameter at a time by its corresponding uncertainty, keeping the others fixed and repeating the analysis. The estimated internal errors are $\pm$100~K in T$_{\mathrm{eff}}$, $\pm$0.15~dex in $\log$~(g), and $\pm$0.15~km s$^{-1}$ in $v_\mathrm{t}$ for the UVES sample, and $\pm$150~K in T$_{\mathrm{eff}}$, $\pm$0.15~dex in $\log$~(g) for the X-Shooter sample. Because the S/N and atmospheric parameters of our sample stars are very close to each other, we estimated the typical errors considering a single reference star. Table~\ref{Tab:errors} lists the effects of these changes on the derived abundances for fnx\_06\_019 in the UVES sample, and Table~\ref{Tab:errors2} lists the effects of these changes on the derived abundances for carLG04c\_0008 in the X-Shooter sample.

{\em Uncertainties due to EWs or spectral fitting.}
The uncertainties on the individual EW measurements ($\delta_{EWi}$)  are provided by {\tt DAOSPEC} (see Table~\ref{Tab:lines}) and are computed according to the formula \citep{Stetson2008} 

\begin{equation}
\delta_{EWi} = \sqrt{\sum_{p}^{}\left(\delta I_p\right)^2 \left(\frac{\partial EW}{\partial I_p}\right)^2+ \sum_{p}^{} \left(\delta I_{C_p}\right)^2 \left(\frac{\partial EW}{\partial I_{C_p}}\right)^2}
,\end{equation}

where $I_p$ and $\delta I_p$ are respectively the intensity of the observed line profile at pixel $p$ and its uncertainty, and $I_{C_p}$ and $\delta I_{C_p}$ are the intensity and uncertainty of the corresponding continuum. The uncertainties on the intensities are estimated from the scatter of the residuals that remain after subtraction of the fitted line (or lines in the case of blends). The corresponding uncertainties $\sigma_{EWi}$ on individual line abundances are propagated by {\tt Turbospectrum}.
This is a lower limit to the real EW error because systematic errors, such as the continuum placement, are not accounted for.

In order to account for additional sources of error, we quadratically added a
5\% error to the EW uncertainty so that no EW has an error smaller than 5\%.
For the abundances derived by spectral synthesis (e.g., strong lines, hyperfine structure, or carbon from the G band), 
the uncertainties were visually estimated by gradually changing the parameters
of the synthesis until the deviation from the observed line became noticeable. 
The abundance uncertainty for an element X due to the individual EW uncertainties ($\sigma_{EWi}$ propagated from $\delta_{EWi}$) are computed as  
 \begin{equation}
     \sigma_{EW}(X) = \sqrt{\dfrac{N_X}{\sum_i 1/\sigma_{EWi}^{2}}}
 ,\end{equation}
where $N_X$ represents the number of lines measured for element X. The dispersion $\sigma_X$ around the mean abundance of an element X measured from several lines is computed as 
 \begin{equation}
 \sigma_X = \sqrt{\dfrac{\sum_i(\epsilon_i - \overline{\epsilon})^2}{N_X-1}}
 ,\end{equation}
where $\epsilon$ stands for the logarithmic abundance.

The final error on the elemental abundances is defined as $\sigma_{fin}$~=~max($\sigma_{EW}$(X), $\sigma_X / \sqrt{N_X}$, $\sigma_{Fe} / \sqrt{N_X}$). As a consequence, no element X can have an estimated uncertainty $\sigma_X$~<~$\sigma_{Fe}$; this is particularly important for species for which the abundances are derived on very few lines.

\begin{table}[ht]
\centering
\caption{Changes in the mean abundances $\Delta$[X/H] caused by a $\pm 100$~K change in $T_{\mathrm{eff}}$, a $\pm$0.15~dex change in $\log$ (g), and a $\pm$0.15~km s$^{-1}$ change in $v_\mathrm{t}$ for the UVES star fnx\_06\_019.}
\resizebox{\linewidth}{!}{
\begin{tabular}{lcccccc}

\hline\hline\Tstrut
 & \multicolumn{6}{c}{$\delta$log$\epsilon$(X)} \\
\hline\Tstrut
El. & $+\Delta$T$_\mathrm{eff}$ & $-\Delta$T$_\mathrm{eff}$ & $+\Delta\log$g & $-\Delta\log$g & $+\Delta$v$_\mathrm{t}$ & $-\Delta$v$_\mathrm{t}$ \\
\Tstrut
 & \multicolumn{6}{c}{ fnx\_06\_019 ( 4280 0.68 1.8 -2.92 )} \\
\hline\Tstrut
\ion{Fe}{I}  & $+0.14$ & $-0.15$ & $+0.00$ & $+0.00$ & $-0.03$ & $+0.03$ \\
\ion{Fe}{II} & $-0.02$ & $+0.04$ & $+0.05$ & $-0.05$ & $+0.02$ & $+0.02$ \\
\ion{ C}{I}  & $+0.13$ & $-0.08$ & $+0.00$ & $+0.01$ & $+0.00$ & $+0.00$ \\
\ion{ O}{I}  & $+0.06$ & $-0.03$ & $+0.06$ & $-0.05$ & $+0.01$ & $+0.01$ \\
\ion{Mg}{I}  & $+0.09$ & $-0.10$ & $-0.02$ & $+0.02$ & $+0.04$ & $+0.04$ \\
\ion{Ca}{I}  & $+0.11$ & $-0.13$ & $-0.02$ & $+0.01$ & $-0.03$ & $+0.02$ \\
\ion{Sc}{II} & $+0.03$ & $-0.01$ & $+0.05$ & $-0.05$ & $+0.03$ & $+0.03$ \\
\ion{Ti}{I}  & $+0.27$ & $-0.26$ & $+0.02$ & $-0.03$ & $+0.03$ & $+0.03$ \\
\ion{Ti}{II} & $+0.02$ & $+0.00$ & $+0.05$ & $-0.04$ & $+0.05$ & $+0.05$ \\
\ion{Cr}{I}  & $+0.20$ & $-0.20$ & $+0.00$ & $+0.00$ & $-0.02$ & $+0.02$ \\
\ion{Co}{I}  & $+0.22$ & $-0.18$ & $+0.01$ & $-0.02$ & $-0.13$ & $+0.15$ \\
\ion{Ni}{I}  & $+0.18$ & $-0.16$ & $+0.01$ & $+0.00$ & $+0.03$ & $+0.03$ \\
\ion{Sr}{II} & $+0.00$ & $+0.00$ & $+0.00$ & $+0.00$ & $+0.00$ & $+0.00$ \\
\ion{ Y}{II} & $+0.05$ & $-0.02$ & $+0.05$ & $-0.05$ & $+0.02$ & $+0.02$ \\
\ion{Ba}{II} & $+0.06$ & $-0.04$ & $+0.06$ & $-0.05$ & $-0.05$ & $+0.07$ \\
\hline

\hline
\end{tabular}}
\label{Tab:errors}
\end{table}

\begin{table}[ht]
\centering
\caption{Changes in the mean abundances $\Delta$[X/H] caused by a $\pm 150$~K change in $T_{\mathrm{eff}}$ with consistent changes in photometric $\log$ (g) and $v_\mathrm{t}$, and by a change of $\pm$0.15~dex in $\log$ (g) only, for the X-Shooter star carLG04c\_0008.}
\resizebox{\linewidth}{!}{
\begin{tabular}{lcccc}

\hline\hline\Tstrut
 & \multicolumn{4}{c}{$\delta$log$\epsilon$(X)} \\
\hline\Tstrut
El. & $+\Delta$T$_\mathrm{eff}$($\log$g,v$_\mathrm{t}$) & $-\Delta$T$_\mathrm{eff}$($\log$g,v$_\mathrm{t}$) & $+\Delta\log$g & $-\Delta\log$g \\
\Tstrut
 & \multicolumn{4}{c}{ carLG04c\_0008 ( 4518 1.08 -3.05 1.78 )} \\
\hline\Tstrut
\ion{Fe}{I}  & $+0.14$ & $-0.23$ & $-0.01$ & $+0.01$ \\
\ion{Fe}{II} & $+0.03$ & $-0.01$ & $+0.06$ & $-0.05$ \\
\ion{ C}{I}  & $+0.22$ & $-0.23$ & $-0.01$ & $+0.01$ \\
\ion{Na}{I}  & $+0.31$ & $-0.36$ & $-0.01$ & $+0.00$ \\
\ion{Mg}{I}  & $+0.30$ & $-0.33$ & $-0.03$ & $+0.03$ \\
\ion{Ca}{I}  & $+0.11$ & $-0.14$ & $-0.01$ & $+0.00$ \\
\ion{Ti}{II} & $+0.11$ & $-0.10$ & $+0.06$ & $-0.07$ \\
\ion{Cr}{I}  & $+0.26$ & $-0.27$ & $+0.02$ & $-0.01$ \\
\ion{Co}{I}  & $+0.25$ & $-0.25$ & $+0.02$ & $-0.01$ \\
\ion{Ni}{I}  & $+0.21$ & $-0.25$ & $+0.01$ & $-0.01$ \\
\ion{Sr}{II} & $+0.22$ & $-0.06$ & $+0.14$ & $-0.16$ \\
\ion{Ba}{II} & $+0.16$ & $-0.13$ & $+0.06$ & $-0.05$ \\
\hline

\end{tabular}}
\label{Tab:errors2}
\end{table}

\begin{figure*}[ht]
\centering
\begin{minipage}[t]{.48\textwidth}
    \centering
    \includegraphics[width=1\columnwidth]{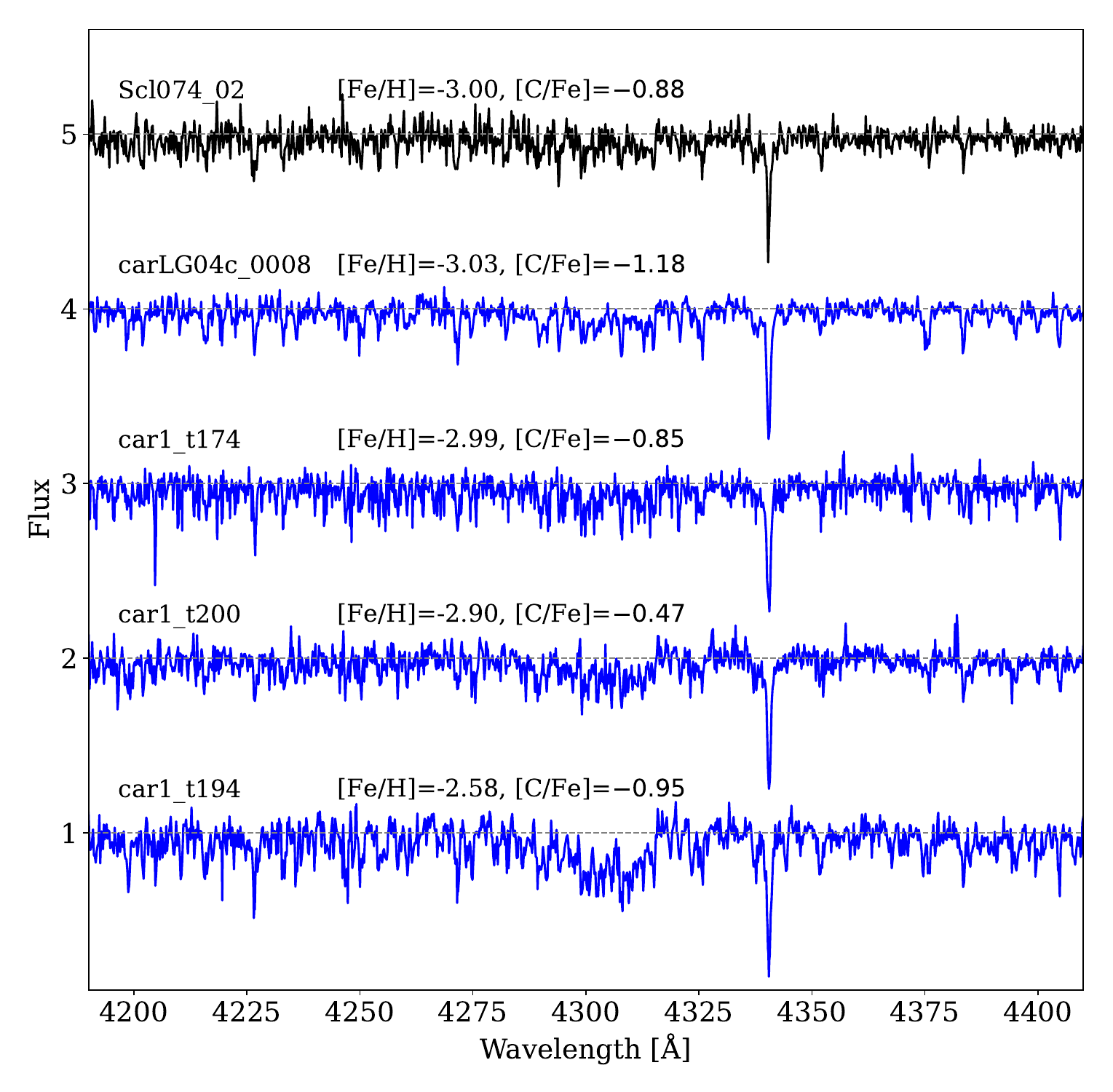}
    \caption{X-Shooter spectra of the four Carina stars (blue) in the region of the CH absorption, from the most metal poor at the  top ([Fe/H]~=~$-3.03$) to the most metal rich at the bottom ([Fe/H]~=~$-2.58$). For comparison   the spectrum of the Sculptor star Scl074\_02 (black, at the top) is shown, with similar atmospheric parameters ([Fe/H]~=~$-3.0$) from \cite{Starkenburg2013}.}
    \label{Fig:CH_spec}
\end{minipage}\hspace{0.5cm}
\begin{minipage}[t]{.48\textwidth}
    \centering
    \includegraphics[width=1\columnwidth]{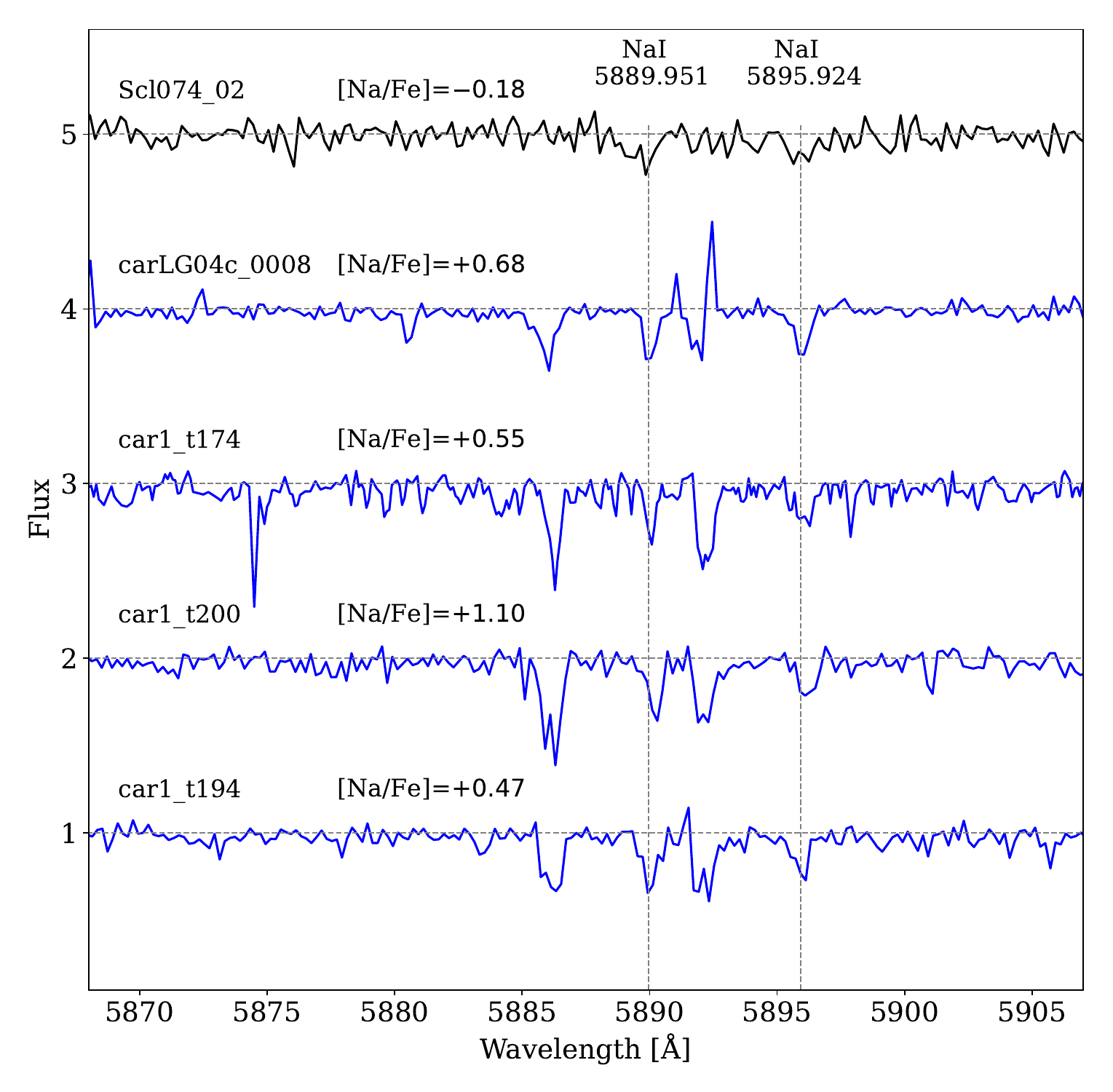}
    \caption{X-Shooter spectra of the four Carina stars (blue) in the region of the \ion{Na}{I} doublet, from the most metal poor at the  top ([Fe/H]~=~$-3.03$) to the most metal rich at the bottom ([Fe/H]~=~$-2.58$), and the spectrum of the Sculptor star Scl074\_02 (black, at the  top) with similar atmospheric parameters ([Fe/H]~=~$-3.0$) from \cite{Starkenburg2013}.}
    \label{Fig:Na_spec}
\end{minipage}
\end{figure*}

\begin{figure*}[ht]
    \centering
    \includegraphics[width=\textwidth]{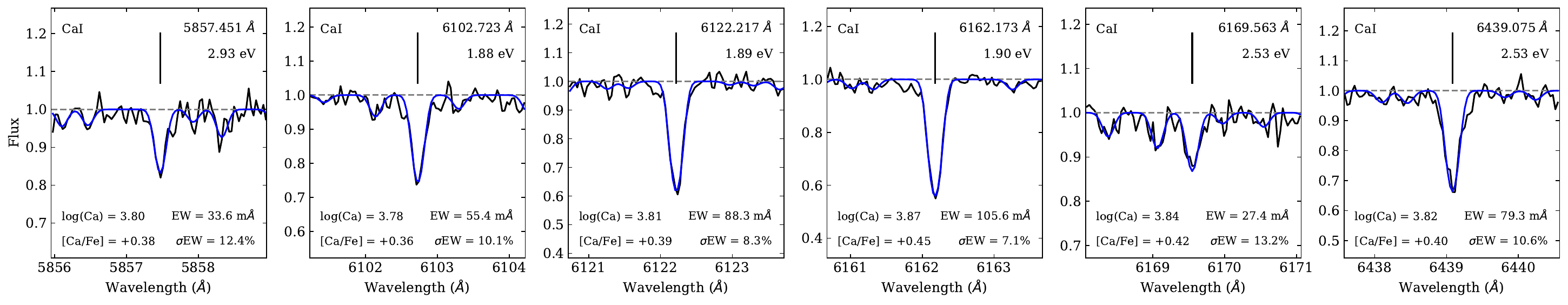}
    \caption{Examples of the \ion{Ca}{I} lines used in the UVES spectrum (black) for the Fornax star fnx\_06\_019. In blue is  the EW fitting from {\tt DAOSPEC}. Atomic data for each line is listed in the panels, along with the derived abundances, which agree very well between different lines.}
    \label{Fig:Ca_spec}
\end{figure*}

%%%%%%%%%%%%%%%%%%%%%%%%%%%%%%%%%%%%%%%%%%%%%%%%%%%%%%%%%%%%%%%%%%%%%%%%%%%%%%%%%%%
\subsection{Specific comments on  abundance determinations}\label{INDABU}
%%%%%%%%%%%%%%%%%%%%%%%%%%%%%%%%%%%%%%%%%%%%%%%%%%%%%%%%%%%%%%%%%%%%%%%%%%%%%%%%%%%

\subsubsection{Carbon}
Carbon abundances were determined by spectral synthesis of the region of the CH molecular G-band.
The carbon abundances of the UVES Fornax sample were determined in the deeper and unblended 4222 -- 4225~\AA\ region, while carbon was determined in a larger range of the CH molecular band, between 4270~\AA\ and 4330~\AA\, for the X-Shooter sample in Carina. We assumed [O/Fe]~=~[Mg/Fe], but since the carbon abundance is low in our sample stars, the exact [O/Fe] assumed has very little effect on our results. The CH molecular band is shown in Fig.~\ref{Fig:CH_spec} for the X-Shooter spectra of our four Carina stars.

\subsubsection{Lithium}
Unfortunately, the Li resonance doublet at 6707~\AA\ was not measurable in any of our stars at our available quality of the UVES or X-Shooter spectra. This was expected, since Li on the surface of RGB stars is typically depleted due to dredge-up and mixing \citep[e.g.,][]{Gratton2000,Lind2009}. None of our target stars are therefore lithium-enhanced giants, which have been found in small fractions in most types of environments, including dwarf galaxies \citep[e.g.,][]{Kirby2012a,Hill2019}.

\subsubsection{Sodium and aluminium}

The sodium was measured from the two 5889.951~\AA\ and 5895.924~\AA\ resonance lines through spectral synthesis.
The X-Shooter spectra of the four Carina target stars are presented in the region of the Na lines and the CH absorption in Fig.~\ref{Fig:Na_spec} and Fig.~\ref{Fig:CH_spec}, respectively. They are compared to a star from \cite{Starkenburg2013} in the Sculptor dSph galaxy, Scl074\_02, with very similar atmospheric parameters (T$_\mathrm{eff}$=4595\,K $\log$~g=1.21 [Fe/H]=$-$3.0 $v_\mathrm{t}$=2.1\,km\,s${^{-1}}$),  but significantly lower Na. 
The carbon abundance is very similar in all stars shown in Fig.~\ref{Fig:CH_spec}, with low [C/Fe]~$\le-0.5$. However, the difference can be clearly seen in the sodium region (Fig.~\ref{Fig:Na_spec}), where the sodium lines appear stronger in Carina than in  Sculptor.
We note that interstellar Na lines are observed at different radial velocities and are unlikely to be contaminating the measured lines.

Two \ion{Al}{I} lines are detected in the UVES spectra, but they are in a noisy part of the spectrum and fall very close to the strong \ion{Ca}{II} H \& K absorption doublet. Furthermore, the continuum level is hard to determine in this region, and the derived abundances strongly depend on it. Because of these difficulties, Al was only derived for one star, fnx\_06\_019 , based on one line at 3961.52~\AA.

\subsubsection{\texorpdfstring{$\alpha$}{a}-elements}

    $\mathit{Magnesium}$. The UVES Mg abundances are based on three lines. Two of them are rather strong (5172.684 and 5183.604~\AA), with EW~>~100~m\AA\ and have non-Gaussian line profiles. The abundances of these lines are not consistent with the weaker 5528.405~\AA\ line. For this reason, we derived the Mg abundance through spectral synthesis, after which all lines had consistent abundances. Three additional \ion{Mg}{i} lines (4167.271, 4351.906, and 5711.088~\AA) were detected in our spectra, but were discarded because   too noisy, strongly blended, and too weak, respectively.
    The Mg abundances from the X-Shooter sample were obtained through spectral synthesis in two 20~\AA\ windows, one centered on the 5172.684~\AA\ line, taking into account the blends at the X-Shooter resolution, and one centered on the 5183.604~\AA\ line. 

    $\mathit{Silicon}$. One \ion{Si}{I} line was detected in our UVES spectra, at 4102.936~\AA, but in a noisy part of the spectrum;   it falls close  to the strong H$\delta$ absorption line. However, we were able to derive a Si abundance in both Fornax stars based on this one line.
    
    $\mathit{Calcium}$. Three \ion{Ca}{I} lines were used in the X-Shooter spectra for the Carina stars. Calcium was not measurable in the star car1\_t200. In the Fornax sample, in the red part of the UVES spectra ( >5500~\AA\,) 6 and 11 clean \ion{Ca}{I} lines were used, for fnx0579x-1 and fnx\_06\_019, respectively. Examples of the \ion{Ca}{I} lines for fnx\_06\_019 are shown in Fig.~\ref{Fig:Ca_spec}.
    
    $\mathit{Titanium}$. The \ion{Ti}{i} abundances are based on 10$-$11 lines, all giving consistent abundance values from their EW. The \ion{Ti}{ii} abundances are based on 8$-$14 lines. They are slightly more scattered as many of them are rather strong.
    The mean abundances of \ion{Ti}{i} and \ion{Ti}{ii} are different by $\Delta$(\ion{Ti}{ii}$-$\ion{Ti}{i})~=~+0.50. This is explained by the fact that \ion{Ti}{ii} is less sensitive to NLTE effects than its neutral state. Thus, following \cite{Jablonka2015}, for the purpose of our discussion we adopted the \ion{Ti}{ii} abundances as the most representative of the Titanium content in our stars.

\subsubsection{Iron-peak elements}

    $\mathit{Chromium}$. Cr abundances were derived from 4 to 5 \ion{Cr}{I} lines in the red part of the UVES spectra; all of them give consistent results from their EWs. Five extra lines were detected in our spectra (4254.352, 4274.812, 4289.73, 5206.023, and 5208.409~\AA), but they were rejected for being too strong (>110~m\AA) or too noisy.
    In the X-Shooter spectra only the strongest $\lambda$5206.023 and 5208.409~\AA\ lines were accessible. Cr abundances were obtained from a single spectral synthesis in a 20~\AA\ wavelength range covering the two lines and taking into account the blends at this resolution.
    
    $\mathit{Manganese}$. Mn abundances rely on the three \ion{Mn}{I} 4030.75~\AA,\ 4033.06~\AA,\ and 4034.48~\AA\ lines. These were synthesized taking into account their HFS components and give consistent abundance results.
    The Mn lines 4041.35~\AA\ and 4823.52~\AA\ are also present in our spectra, but they are weak ($\sim$30~m\AA) and too   affected by   noise, and were thus discarded.
    In the case of X-Shooter spectra, a single spectral synthesis was done in a 20~\AA\ window centered on the Mn triplet.

\subsubsection{Neutron-capture elements}

    $\mathit{Strontium}$. Three \ion{Sr}{II} lines are present in our spectra (4077.709, 4161.792, and 4215.519~\AA).
    While the 4215.519~\AA\ line was the most reliable in the UVES spectra, the 4077.709~\AA\ line was less affected by blends and better suited for the X-Shooter spectra.
    
    $\mathit{Barium}$. Ba abundances were measured from two to four lines (4554.029, 4934.076, 5853.668, 6141.713, and 6496.897~\AA) in the UVES and X-Shooter spectra by spectral synthesis taking into account blends and HFS.

\begin{figure}[ht]
        \centering
        \includegraphics[width=0.75\columnwidth]{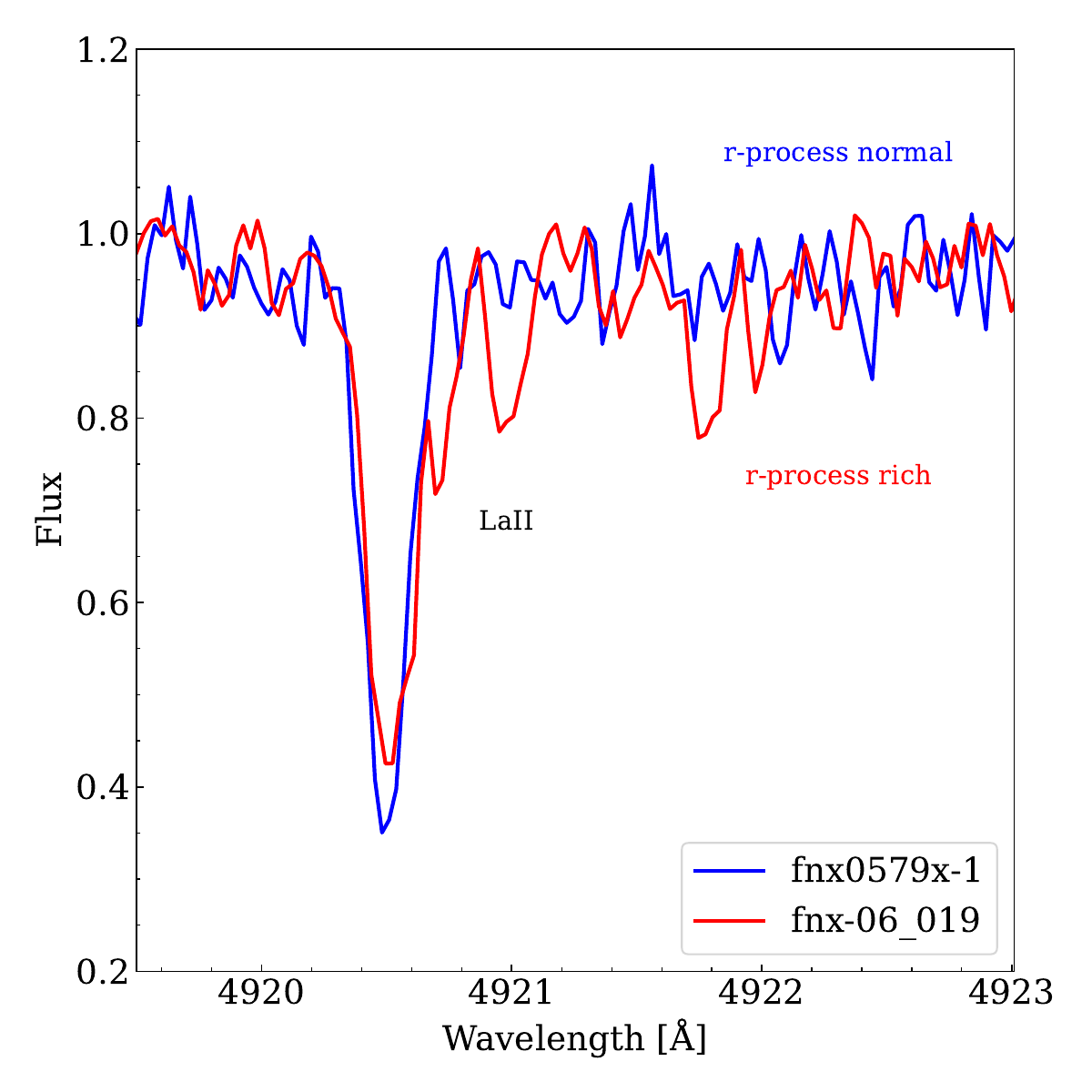}
        \includegraphics[width=0.75\columnwidth]{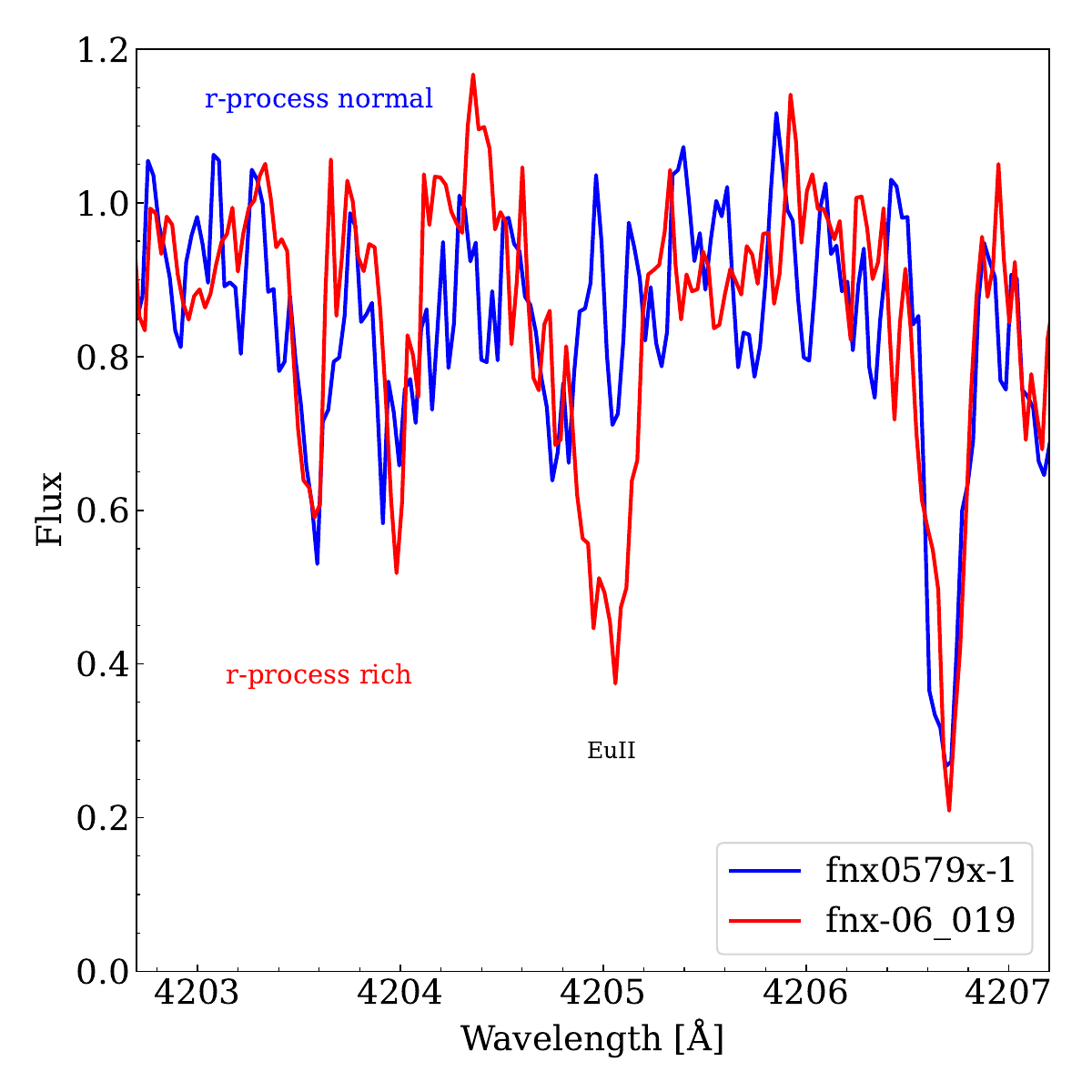}
    \caption{Two UVES spectra in the region of the lanthanum (top) and europium (bottom) lines. In red is the r-process rich star fnx\_06\_019, and in blue the r-process normal star fnx0579x--1.}
    \label{Fig:ncapture-spec}
\end{figure}

Additional neutron-capture elements were measurable in the spectrum of the $r$-process rich star fnx\_06\_019 in Fornax. Figure~\ref{Fig:ncapture-spec} compares two UVES spectra in the region of the lanthanum and europium lines, showing a clear neutron-capture enhancement in fnx\_06\_019 (see also Fig.~\ref{Fig:individual_lines}).

    $\mathit{Lanthanum}$. The La abundance was determined from the $\lambda$4920.98~\AA\ line, which is the reddest available \ion{La}{II} line and the least affected by noise. 
    
    The detection was checked by computing synthetic spectra for the 4077.34, 4086.71, and 4123.23~\AA\ lines using the abundance derived from the 4920.98~\AA\ line (Fig.~\ref{Fig:individual_lines}). All the lines were in good agreement with the adopted abundance. 
    
    $\mathit{Neodymium}$. The Nd abundance was determined from the two clean and unblended \ion{Nd}{II} lines at 4825.48 and 5319.81~\AA , and further confirmed to be consistent with the 4109.45 and 4061.08~\AA\ lines.
    
    $\mathit{Dysprosium}$. The abundance was measured from the \ion{Dy}{II} 4103.31~\AA\ line by spectral synthesis. The 3944.68~\AA\ line is too affected by noise and continuum level uncertainty, while the 4449.7~\AA\ line is too strongly blended to be reliable.

%%%%%%%%%%%%%%%%%%%%%%%%%%%%%%%%%%%%%%%%%%%%%%%%%%%%%%%%%%%%%%%%%%%%%%%%%%%%%%%%%%%
\section{Results}\label{Sec:results}
%%%%%%%%%%%%%%%%%%%%%%%%%%%%%%%%%%%%%%%%%%%%%%%%%%%%%%%%%%%%%%%%%%%%%%%%%%%%%%%%%%%

The measured LTE chemical abundances for our four Carina, and two Fornax stars are listed in Table~\ref{Tab:abundances} and shown in Figs.~\ref{Fig:C}--\ref{Fig:BaEu}. The different element groups are discussed in the following sections.

\subsection{Literature comparison samples}

\begin{figure}[ht]
        \centering
        \includegraphics[width=1.0\columnwidth]{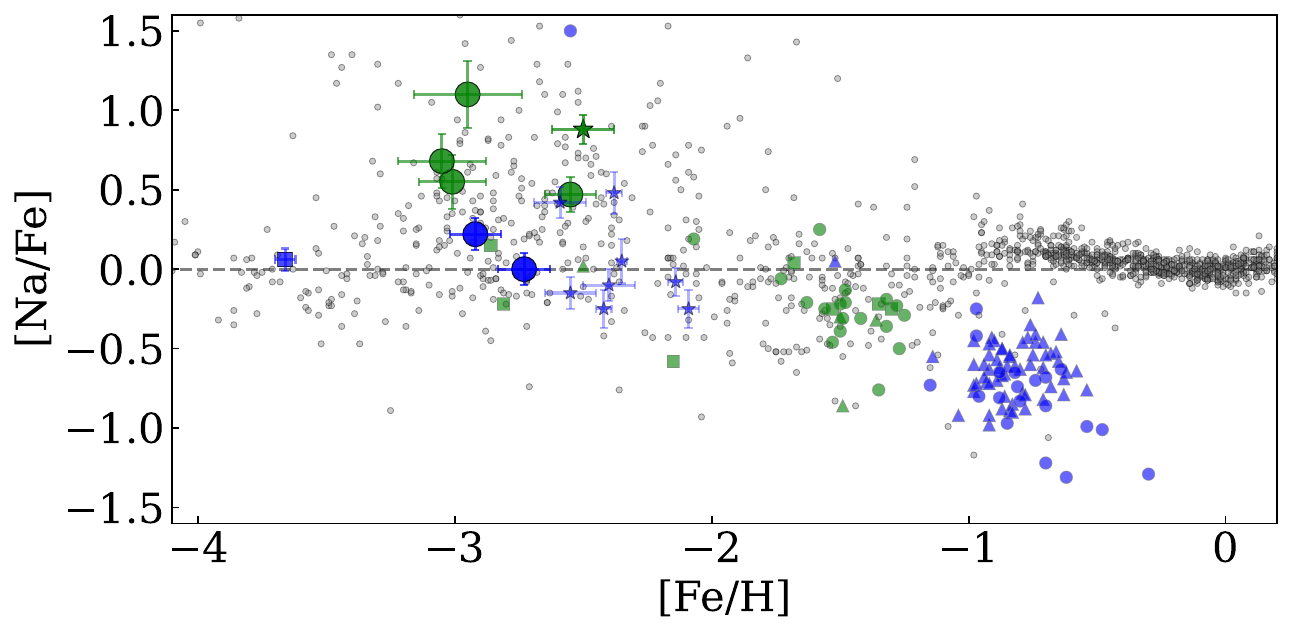}
    \caption{LTE Sodium-to-iron ratios as a function of [Fe/H] are shown for stars in the Fornax and Carina dSph galaxies, and in the MW. Fornax members are in blue: large circles are from this work; small circles are from \cite{Lemasle2014}; small triangles are from \cite{Letarte2010}; small stars are members of Fornax globular clusters 1, 2, and 3 from \cite{Letarte2006}. The square at [Fe/H]~=~--3.66 is an EMP star from \cite{Tafelmeyer2010}. Carina members are in green: large circles are from this work; small triangles are from \cite{Koch2008}; small circles are from \cite{Norris2017}; small squares are from \cite{Venn2012};   the star  is a CEMP-no from \cite{Susmitha2017}. The gray dots are MW stars \citep{Bensby2014,Yong2013,Frebel2010}.}
    \label{Fig:Na}
\end{figure}

In Figures~\ref{Fig:C}--\ref{Fig:BaEu}, the abundances from the Milky Way are shown as gray dots and are compiled from \cite{Placco2014} for carbon; from \cite{Bensby2014}, and \cite{Yong2013} for Na, Mg, Ca, Ti; and from \cite{Frebel2010} for Na. The Milky Way iron-peak elements (Sc, Cr, Mn, Co, Ni, Zn) are from \cite{Bensby2014}, \cite{Yong2013}, and \cite{Frebel2010}; while the abundances of the neutron-capture elements strontium and barium are   from \cite{Roederer2013}. The Milky Way europium abundances are from \cite{Frebel2010}, and \cite{Venn2004}.

Abundances measured in Carina are from \cite{Koch2008}, \cite{Venn2012}, and \cite{Norris2017}. \cite{Norris2017} compiled 63 RGB stars, of which 14 stars were totally new observations, while 18 stars were re-observations and 31 stars were re-analyses of stars initially studied by \cite{Lemasle2012}, \cite{Shetrone2003}, and \cite{Venn2012}. 
For the purpose of homogeneity we show the \cite{Norris2017} results and refer to this paper for a detailed description of the individual samples. The only exception is made for the \cite{Venn2012} sample. Since their work includes more chemical elements than \cite{Norris2017}, we kept the original results of the nine Carina members from \cite{Venn2012}. The Fornax comparison samples are taken from \cite{Letarte2006}, \cite{Letarte2010}, \cite{Lemasle2014}, and \cite{Tafelmeyer2010}.

\subsection{Carbon}

\begin{figure}[ht]
    \centering
    \includegraphics[width=1.0\columnwidth]{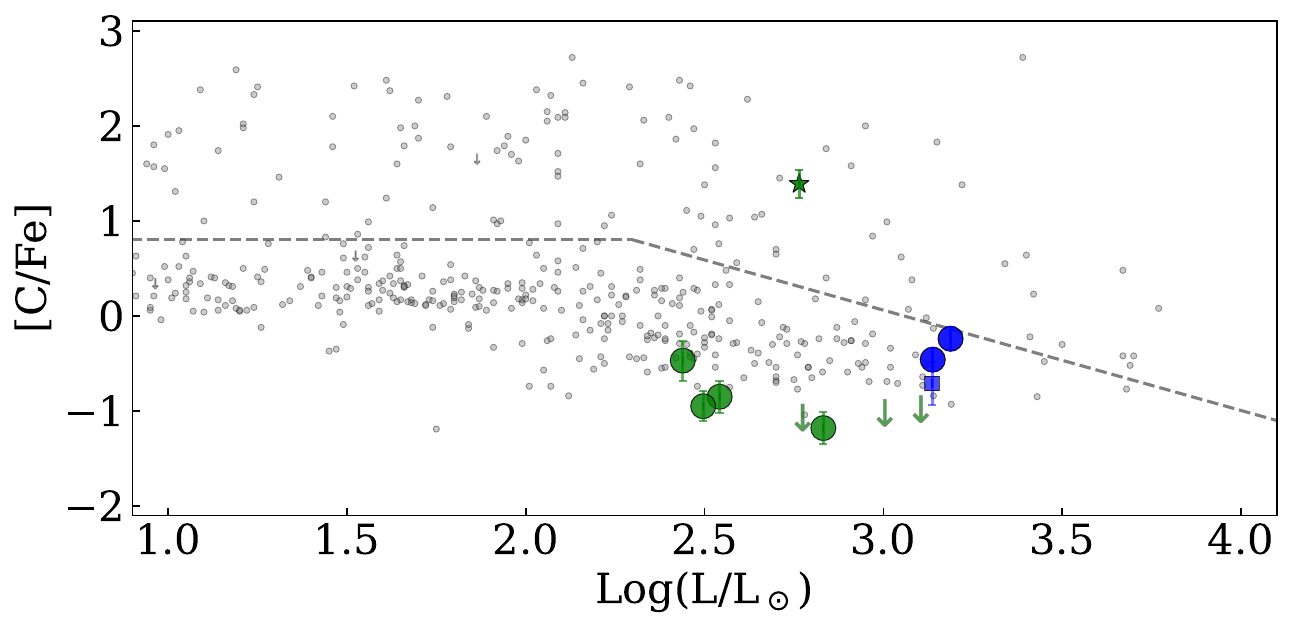}
    \caption{[C/Fe] as a function of $\log$(L/L$_{\odot}$) for Fornax (blue) and Carina (green).
    The circles are data from this work. The green upper limits are from \cite{Venn2012}, and the green star  is a CEMP-no star from \cite{Susmitha2017}. The blue square is the Fornax member from \cite{Tafelmeyer2010}. The gray dots are MW halo stars \citep{Placco2014}. The dotted line traces the criterion of \citet{Aoki2007} to define C-enhanced stars, which takes into account the depletion of carbon along the RGB. Only stars with $\rm[Fe/H]<-2.5$ are included.}
    \label{Fig:C}
\end{figure}

Figure~\ref{Fig:C} presents the carbon abundances measured in Fornax and Carina at [Fe/H]~<~$-$2.5, compared to the  literature data. The four Carina stars analyzed in this paper all present very low carbon levels, similar to the upper limits derived in the Carina sample of \cite{Venn2012}. The Carina [C/Fe] distribution is located at the lower edge of the Milky Way abundances, making Carina noticeably C-poor (see Fig.~\ref{Fig:CH_spec}). However, we note that two CEMP-no stars have been identified in Carina, first in \cite{Susmitha2017} and more recently in \citet{HansenT2023}.

In Fornax, our two stars and the one star previously studied by \citet{Tafelmeyer2010} have similar carbon-normal abundances. On average the Fornax stars are somewhat higher in C compared to Carina, especially when corrections for evolutionary status are taken into account. 

We note that none of our six new stars is enhanced in carbon, while in the Milky Way the typical fraction of CEMP-no stars is $\approx40\%$ at $\rm[Fe/H]<-3$ \citep{Placco2014}. This adds to previous results, showing that dSph galaxies are poor in CEMP-no stars relative to the Milky Way and the UFDs \citep{Starkenburg2013, Skuladottir2015, Skuladottir2021,Skuladottir2023a, Jablonka2015, Simon2015,Kirby2015,HansenC2018,Lucchesi2020}, for a more detailed discussion on this see Section~\ref{Sec:Conclusion}.

\subsection{Light element: Sodium}\label{light}

The LTE abundances of sodium are presented in Fig.~\ref{Fig:Na}. The Milky Way has a very large scatter in [Na/Fe] at low [Fe/H]~<~$-2$, and our new data points fall within this range.  Fornax shows a near solar $\rm[Na/Fe]\approx0$; instead,  all four stars in Carina have high $\rm[Na/Fe]\gtrsim+0.5$. Since direct comparison between the Carina (X-Shooter) and Fornax (UVES) spectra is not meaningful, in Fig.~\ref{Fig:Na_spec} we plot the Na lines of the Carina stars together with an X-Shooter spectrum of a Na-poor star in Sculptor. This comparison clearly shows the Carina stars to be Na rich.

Non-LTE effects can play an important role in the derived sodium abundances. The resonance lines, 5889.951~\AA\ and 5895.924~\AA, are especially sensitive. Very limited NLTE abundances are available in the literature. The apparent large dispersion at [Fe/H]~<~$-2$ could probably be significantly reduced if NLTE corrections were applied. 
We used the {\tt INSPECT} database\footnote{\url{http://www.inspect-stars.com/}} to compute corrections for our stars to investigate whether the [Na/Fe] difference between Fornax and Carina would remain (see Table~\ref{Tab:NLTE_corrections}). 
The correction depends both on the measured LTE sodium abundances and on the metallicity and the stellar atmospheric parameters. 
The corrections for Carina were $\langle \Delta$[Na/Fe]$_\text{NLTE}\rangle=-0.33$ and for Fornax $\langle \Delta$[Na/Fe]$_\text{NLTE}\rangle=-0.16$. However, the Carina NLTE abundances are still higher, with $\langle $[Na/Fe]$_\text{NLTE}\rangle=+0.37\pm0.17$;  for Fornax $\langle $[Na/Fe]$_\text{NLTE}\rangle=-0.06\pm0.14$. 

The interpretation of this Na-enhancement is not straightforward. \citet{Norris2013} found that CEMP-no stars were likely to be enhanced in Na; however, the Carina stars are depleted in carbon (see the previous section, and Fig.~\ref{Fig:C}). The origin of this Na enhancement, compared to other dSph galaxies,  therefore remains unclear. We note however that there are two stars from \citet{Venn2012} with lower Na, so it seems that there is a significant scatter of Na in Carina, similar to the Milky Way.

\begin{figure}
        \centering
        \includegraphics[width=1.0\columnwidth]{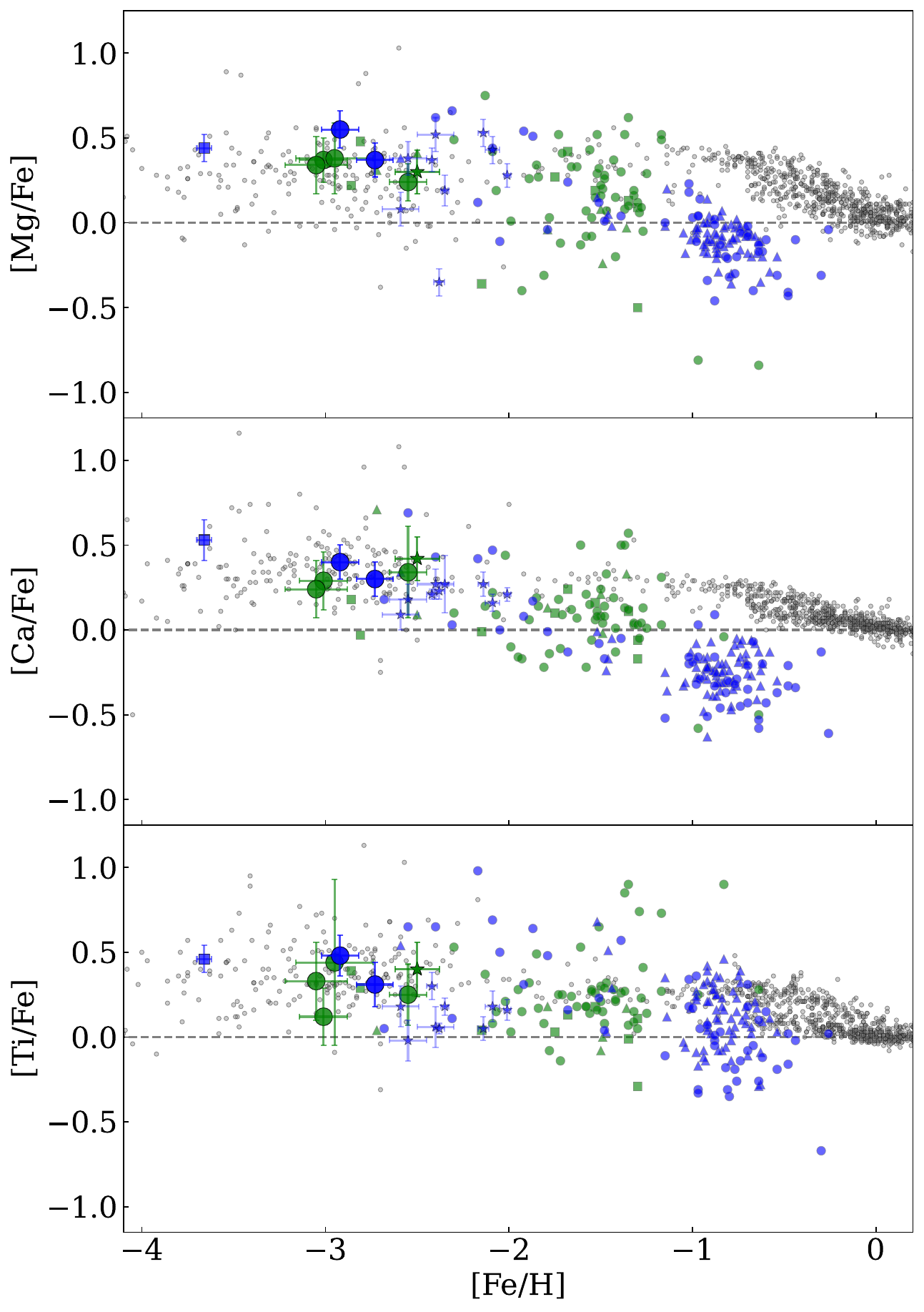}
        \caption{Abundance ratios for the $\alpha$-elements Mg, Ca, and Ti (from top to bottom) as a function of [Fe/H]. The symbols are the same as in Fig.~\ref{Fig:Na}.}
        \label{Fig:alpha}
\end{figure}

\subsection{ \texorpdfstring{$\alpha$}{a}-elements}\label{alpha}

The chemical abundances of the $\alpha$-elements Mg and Ca, together with Ti, are shown in Fig.~\ref{Fig:alpha}. All three elements show a very similar trend of enhanced $\rm[\alpha/Fe]\approx+0.4$ at $\rm[Fe/H]<-2.5$ in all our literature samples, and in \cite{Hendricks2014a} who focuses on determining the position of the $\alpha$-knee in Fornax. Our results are in perfect agreement with the Milky Way, and other dwarf galaxies, both the dSph galaxies \citep[e.g.,][]{Tafelmeyer2010,Lucchesi2020,Theler2020} and the  UFDs \citep[e.g.,][]{Simon2019}.

The Carina dSph galaxy has a complex star formation history \citep[e.g.,][]{Tolstoy2009,deBoer2014}, and the scatter of [Mg/Fe] at intermediate metallicities, $\rm -2.5<[Fe/H]<-1$, is quite large, $\sigma=0.30$~dex. However, the scatter seems to be reduced at $\rm [Fe/H]\leq-2.5$, where all stars are consistent with the same [Mg/Fe] value, $\sigma=0.08$~dex. It is therefore possible that these low metallicities probe only the first star formation burst in Carina.

\begin{figure*}[ht]
    \includegraphics[width=\textwidth]{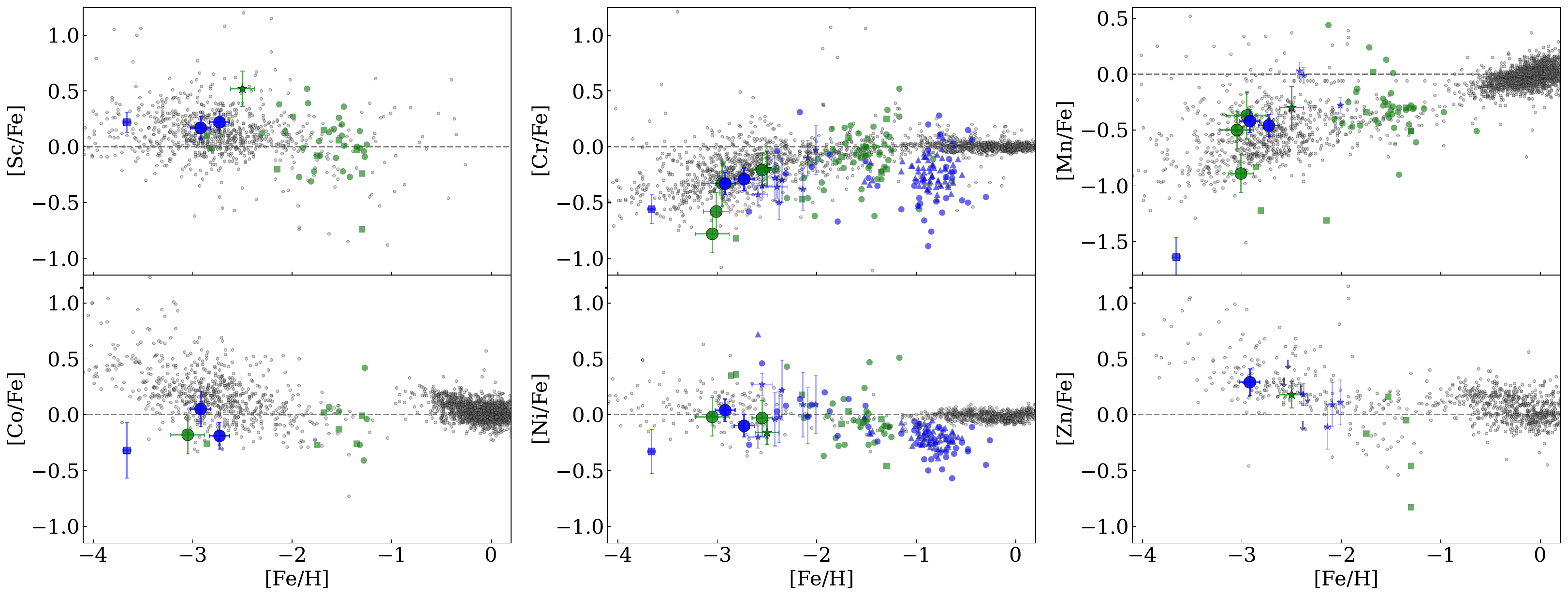}
     \caption{Iron-peak elements as a function of [Fe/H]. From left to right, top to bottom: [Sc/Fe], [Cr/Fe], [Mn/Fe], [Co/Fe], [Ni/Fe], and [Zn/Fe] for metal-poor stars in Fornax, Carina, and MW. The symbols are the same as in Fig.~\ref{Fig:Na}. The stars studied in this paper are the large circles. The MW data are from \cite{Bensby2014}, \cite{Yong2013}, and \cite{Frebel2010}.}
    \label{Fig:FePeak}
\end{figure*}

\subsection{Iron-peak elements}

Figure~\ref{Fig:FePeak} presents Sc, Cr, Mn, Co, Ni, and Zn trends with Fe in Carina and Fornax, compared with the Milky Way halo and disk populations.
The element Sc could  be determined only in our Fornax sample, which had high-resolution UVES spectra. The production of Sc is dominated by core-collapse supernovae (ccSNe; \citealt{Woosley2002, Battistini2015}), and therefore, as expected, [Sc/Fe] is at the same level as the $\alpha$-elements seen in Fig.~\ref{Fig:alpha}.

For both Fornax and Carina, Cr and Mn closely follow the Milky Way trends at $\rm[Fe/H]<-2.5$, as derived from 1D LTE methods.  NLTE calculations for the neutral species of these elements show an overionization, leading to weakened lines and positive NLTE abundance corrections \citep{Bergemann2019}. Therefore, the increasing trends of [Cr/Fe] and [Mn/Fe] with [Fe/H] might be artificial. In the case of Co, our measurements are lower than the average trend in the Milky Way. Similarly, the Co abundances in the Sculptor dSph galaxy are low at $\rm[Fe/H]<-2.5$ \citep{Skuladottir2023b}.

Nickel can be produced in ccSNe, and in thermonuclear  Type Ia supernovae \citep[SNIa; e.g.,][]{Jerkstrand2018}. Before SNIa start to dominate, $\rm[Fe/H]<-2$, our sample definitely set the level at solar value, $\rm[Ni/Fe]\approx0$, for both Carina and Fornax, implying that the global production of nickel follows that of iron in core-collapse supernovae. This is in excellent agreement with the Sculptor dSph galaxy and UFDs \citep{Skuladottir2023b}.
However, subsolar [Ni/Fe] has been noted in the Fornax population at $\rm[Fe/H]>-1.2$ \citep{Letarte2010,Lemasle2014}, and in Carina at $\rm[Fe/H]>-1.5$ \citep{Norris2017}. This clearly corresponds to a stage of the galaxy chemical evolution when the ejecta of SNIa dominate the composition of the interstellar medium. Similar results are found in other dSph galaxies \citep{Hill2019,Kirby2019,Theler2020}, as well as accreted dwarf galaxies \citep[e.g.,][]{NissenSchuster2010}.

There is a known increase in [Zn/Fe] toward low [Fe/H] \citep[e.g.,][]{Cayrel2004} (see Fig.~\ref{Fig:FePeak}, bottom right). The NLTE corrections of Zn \citep{Takeda2005} are small in the metal-poor regime and do not fully explain this trend, and it remains poorly understood. We were only able to measure Zn for the star fnx\_06\_019, which falls on the Milky Way halo trend at $\rm[Zn/Fe]=+0.29$, comparable to the level of the [$\alpha$/Fe] plateau. This was also observed in two Sextans EMP stars \citep{Lucchesi2020}, two stars in Ursa Minor \citep{Cohen2010}, and in Sculptor \citep{Skuladottir2017,Skuladottir2023b}, highlighting the role of ccSNe in the production of zinc in the early stage of galaxy evolution.

\begin{figure}[ht]
    \centering
    \includegraphics[width=1.0\columnwidth]{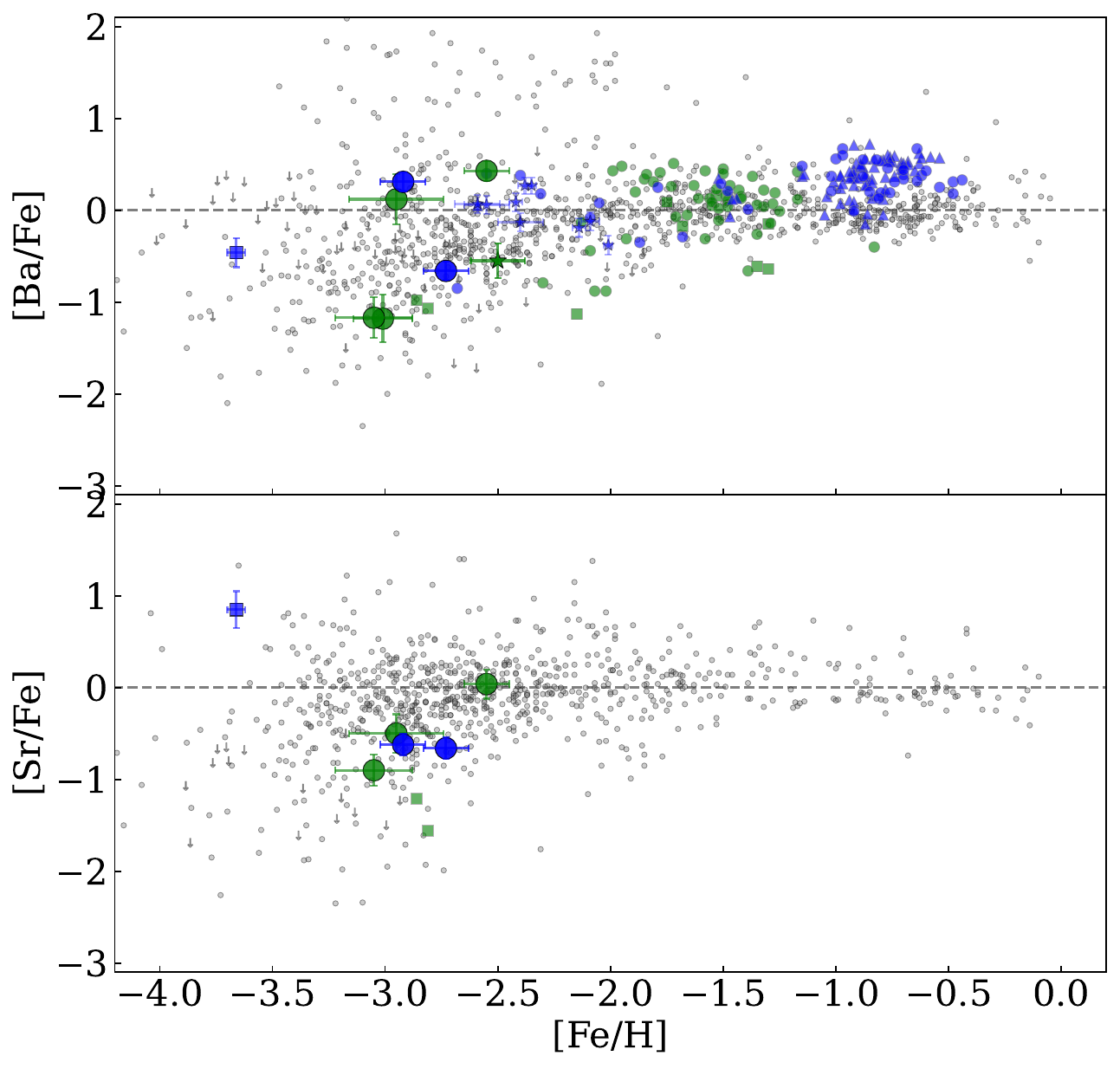}
    \caption{Neutron-capture elements. Shown are the barium-to-iron ratio (top) and strontium-to-iron ratio (bottom) as a function of [Fe/H] in Fornax (blue) and Carina (green), compared to MW stars (gray) from \cite{Roederer2013}. The symbols are the same as in Fig.~\ref{Fig:Na}; the large circles represent the new sample analyzed here.}
    \label{Fig:ncapture}
\end{figure}

\begin{figure}[ht]
    \centering
    \includegraphics[width=1.0\columnwidth]{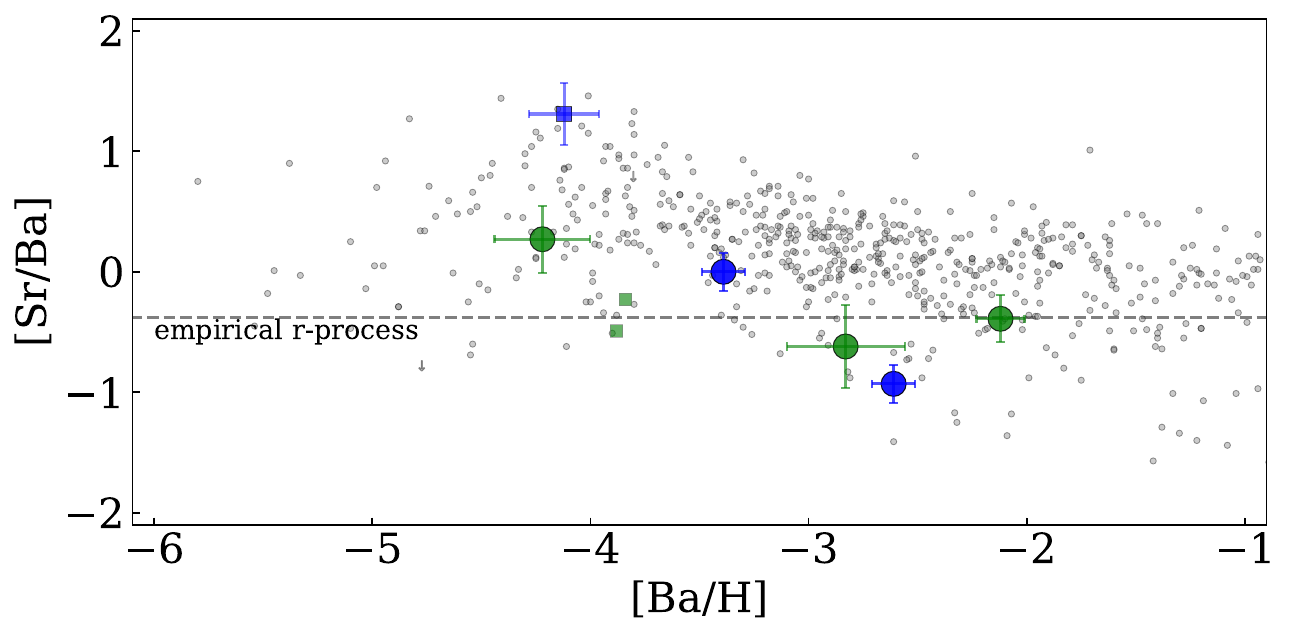}
    \caption{Barium-to-strontium ratio as a function of [Ba/H]. The references are the same as in Fig.~\ref{Fig:ncapture}. The empirical r-process limit is shown as a dashed line \citep{Mashonkina2017}.}
    \label{Fig:SrBa}
\end{figure}

\begin{figure}[ht]
    \centering
    \includegraphics[width=1.0\columnwidth]{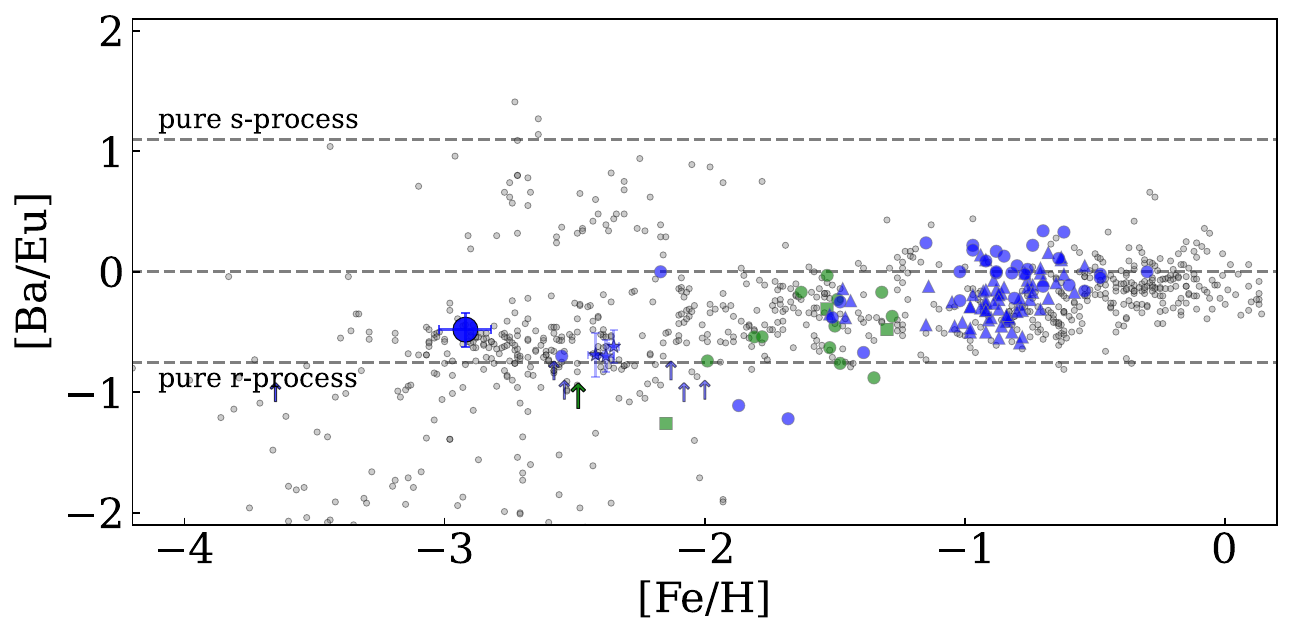}
    \caption{Barium-to-europium ratio as a function of [Fe/H]. The symbols are the same as in Fig.~\ref{Fig:ncapture}. The MW stars are from \cite{Frebel2010}, and \cite{Venn2004}. The limits of pure s-process and r-process are shown as dashed lines \citep{Mashonkina2017}.}
    \label{Fig:BaEu}
\end{figure}

\subsection{Neutron-capture elements}

Figure~\ref{Fig:ncapture} shows our measurements of [Sr/Fe] and [Ba/Fe] with [Fe/H]. The Ba and Sr abundances of our sample stars are consistent with the large spread in the Milky Way. 
The abundance ratio of [Sr/Ba] with [Ba/H] is presented in Fig.~\ref{Fig:SrBa}. 
The [Sr/Ba] ratios of the Fornax stars are compatible with those of the Milky Way. However, these ratios in the Carina dSph galaxy are somewhat lower.
\citet{Mashonkina2017} identified two populations in [Sr/Ba], one with a trend  similar to that found in the Milky Way (and Fornax), increasing toward low [Ba/H], and the other of near constant [Sr/Ba], similar to the empirical r-process ratio. The Carina abundances   seem to fall   between these two populations.

Only in the r-I star, fnx\_06\_019, were we able to measure [Eu/Fe]. Its [Ba/Eu]~=~$-0.5$ value sits just above the pure r-process ratio, as is shown in Fig.~\ref{Fig:BaEu}. This star, therefore, likely also shows some minor s-process contribution, which is   evident from its rather high $\rm[Sr/Ba]=+0.5$ (see Fig.~\ref{Fig:SrBa}). Not only is this star rich in Eu, but it is also enhanced in La, Nd, and Dy, as is evident from the very strong lines shown in Fig.~\ref{Fig:individual_lines} and listed Table~\ref{Tab:abundances}. However, we note that the light neutron-capture elements, Sr, Y, and Zr, are not enhanced:  $\rm[Sr/Fe]=-0.62$, $\rm[Y/Fe]=-0.15$,  $\rm[Zr/Fe]=+0.22$. The elements Y, Sr, and Zr have a similar origin, and their ratios are observed to be approximately constant in the Milky Way halo \citep{Francois2007}. 

This is the first r-I star identified in a dSph galaxy at such low [Fe/H]. The r-I and r-II stars discovered in dSph galaxies to date seem to cover the more metal-rich end of the metallicity distribution compared to halo r-I stars \citep[e.g.,][]{Shetrone2001,Aoki2007,Cohen2009,Hansen2018,Reichert2020}.

%%%%%%%%%%%%%%%%%%%%%%%%%%%%%%%%%%%%%%%%%%%%%%%%%%%%%%%%%%%%%%%%%%%%%%%%%%%%%%%%%%%
\section{Discussion and conclusions}\label{Sec:Conclusion}
%%%%%%%%%%%%%%%%%%%%%%%%%%%%%%%%%%%%%%%%%%%%%%%%%%%%%%%%%%%%%%%%%%%%%%%%%%%%%%%%%%%

Here we follow-up some of the first EMP candidates in the Fornax and Carina dSph galaxies to populate the as-yet-uncovered metallicity range $\rm -3.1\le[Fe/H]\le -2.5$. It is now clear that regardless of the subsequent evolution of the local classical dwarf galaxies, which harbor very different star formation histories, the first generations of stars formed in very similar way in these systems. Almost all the chemical elements follow the same trend with (low) metallicity, and match the known relations for our Galaxy.

\begin{figure}[ht]
        \centering
        \includegraphics[width=1.0\columnwidth]{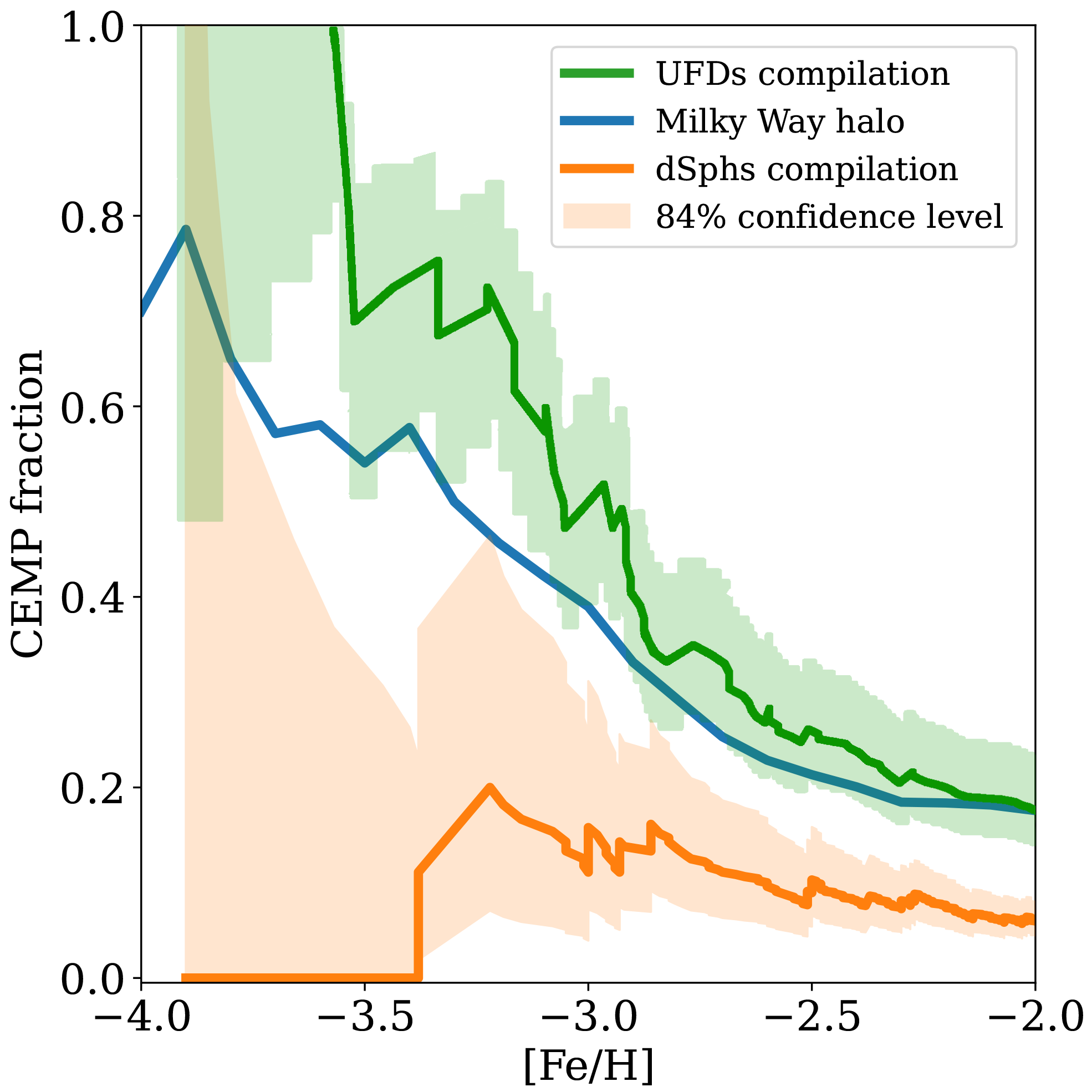}
    \caption{Cumulative fraction of CEMP-no stars ([C/Fe]$\ge+0.7$) observed in dSph galaxies (orange solid line) with the associated 1$\sigma$ 84.13\% confidence level \citep{Gehrels1986} (orange shaded area). Fraction in the MW halo (blue solid line) is from \cite{Placco2014}, 
    while the fraction computed by \cite{Ji2020} in UFDs is represented by the green solid line with its associated uncertainty (green shaded area).
    The carbon abundances in dSph galaxies are corrected for internal mixing \citep{Placco2014} and are compiled from these works:  \cite{Lucchesi2020,Jablonka2015,Tafelmeyer2010,Skuladottir2015,Kirby2015,Venn2012,Starkenburg2013,Susmitha2017,HansenC2018,Yoon2020,Kirby2012,Skuladottir2023b}.}
    \label{Fig:C_frac}
\end{figure}

However, there are clear differences in the [C/Fe] abundances of dSph galaxies, compared to UFDs and the Milky Way. 
In Fig.~\ref{Fig:C_frac} we compare the cumulative fraction of CEMP-no stars in dSph galaxies to that of the Milky Way,  as \citet{Ji2020} did for UFDs. 
We used all the available literature data, corrected for carbon depletion due to  stellar evolution \citep{Placco2014},\footnote{\url{https://vplacco.pythonanywhere.com/}} that had  minimal or no selection effects in regards to C-abundance. 
In particular, we were unable to include \citet{HansenT2023}, as they specifically targeted CEMP stars, thus not providing information about the CEMP-no fraction. No CEMP-s stars (i.e. CEMP stars presenting s-process enhancement, [Ba/Fe]~>~0) were included; however, we did include the $\rm[Fe/H]=-3.38$ star in Canes Venatici~I, without a Ba measurement, which \citet{Yoon2020} claimed was a CEMP-no star, using the stars position in the [Fe/H]-A(C) diagram. We note that [C/Fe] in this star is $\gtrsim1$~dex higher than all other dSph stars found in the literature. 
From Fig.~\ref{Fig:C_frac} it is evident that the CEMP-no fraction in dSph galaxies is significantly lower than in the Milky Way, reaching fractions of $\approx6\%$ at $\rm[Fe/H]<-2$, based on 15 CEMP-no stars, out of 251 in total. At the lowest metallicities, $\rm [Fe/H]\leq-3.4$, none of the seven known stars in dSph galaxies is C-enhanced, while the CEMP-no fraction in the Milky Way at these metallicities is $>50\%$. This also implies a discrepancy in the CEMP-no fraction of dSph galaxies and UFDs, since the latter are in agreement with the Milky Way fraction \citep[e.g.,][]{Ji2020}. Our findings therefore confirm what has previously been stated with more limited data (e.g., \citealt[][]{Starkenburg2013,Skuladottir2015,Skuladottir2023b,Lucchesi2020}).

We see differences in the Na abundances of Fornax and Carina, where all   four stars in the Carina dSph galaxy are high in $\rm[Na/Fe]_\text{LTE}\gtrsim+0.5$ (see Fig.~\ref{Fig:Na}). This is puzzling since high Na abundances are typically found in CEMP-no stars \citep[e.g.,][]{Norris2013}, and are predicted by their theoretical yields \citep[e.g.,][]{HegerWoosley2002}, but the Carina stars all have low $\rm[C/Fe]<0$. Furthermore, this is typically not seen in other very metal-poor stars in dSph galaxies \citep{Tafelmeyer2010,Jablonka2015,Skuladottir2023b}.

As we do  in the Milky Way, at low $\rm[Fe/H]\lesssim-2.5$ we find a large scatter of the neutron-capture elements in both the Carina and the Fornax dSph galaxies. 
We report the discovery of a Eu-rich, r-I star in Fornax: $\rm[Eu/Fe]=+0.8$ and $\rm[Eu/Ba]=+0.5$.  This star, fnx\_06\_019, also shows an outstanding enrichment in La, Nd, and Dy ([X/Fe]>+0.5). This is the first such case for a C-normal star at such low metallicity ($\rm[Fe/H]=-2.92$) in a classical dSph galaxy. This points to a prolific r-process event early in the history of the Fornax dSph galaxy, where r-process enhancements were previously only seen at higher [Fe/H]~>~--2 \citep[][]{Letarte2010, Lemasle2014,Reichert2020,Reichert2021}.

The present analysis contributes to the limited understanding of the earliest chemical enrichment in dSph galaxies. To further investigate the range of abundance patterns observed in different systems, it is clear that more observations are needed. Fortunately, the situation will improve drastically with upcoming spectroscopic surveys, such as WEAVE \citep{Jin2023} and 4MOST \citep{deJong2019}, in particular the dedicated dwarf galaxy survey 4DWARFS \citep{Skuladottir2023a}. With large and homogeneous data sets we will be able to identify rare EMP stars, and better constrain the CEMP-no fraction in different galaxies. Furthermore, we will obtain detailed spatial information that will allow us to trace the individual events causing r-process enrichment. Our work serves to highlight that it is fundamental to study metal-poor stars in galaxies of different sizes and star formation histories to get a complete picture of the galaxy formation and first chemical enrichment.

\begin{acknowledgements}

We thank E. Tolstoy for useful comments and suggestions on the manuscript. 
R.L. and \'{A}.S have received funding from the European Research Council (ERC) under the European Union’s Horizon 2020 research and innovation programme (grant agreement No. 804240).
D.M. gratefully acknowledges support from the ANID BASAL projects ACE210002 and FB210003, from Fondecyt Project No. 1220724, and from CNPq Project 350104/2022-0.

\end{acknowledgements}

\bibliographystyle{aa}
\bibliography{main-paper}

\clearpage

%%%%%%%%%%%%%%%%%%%%%%%%%%%%%%%%%%%%%%%%%%%%%%%%%%%%%%%%%%%%%%%%%%%%%%%%%%%%%%%%%%%
\begin{appendix}
%%%%%%%%%%%%%%%%%%%%%%%%%%%%%%%%%%%%%%%%%%%%%%%%%%%%%%%%%%%%%%%%%%%%%%%%%%%%%%%%%%%

\section{Additional tables}

\begin{table}[htp]
\caption{Lines measured in the Fornax UVES spectra. Line parameters, observed EWs, and elemental abundances are provided. 
The EWs in brackets are given only as an estimation of the strength of the line. They should not be used to derive chemical abundances since most of them are blended, have large uncertainties, or derive from a Gaussian profile; the quoted abundances are derived through spectral synthesis for these lines.
}
\resizebox{\linewidth}{!}{
%\resizebox{0.7\linewidth}{!}{
\begin{tabular}{lllr|rrrr}

\hline
\hline\Tstrut
El. & $\lambda$ & $\chi_{ex} $ & log($gf$) & EW [m\AA] & log$\epsilon$(X) & EW [m\AA] & log$\epsilon$(X) \\
 & [\AA] & [eV] &  &\multicolumn{2}{c}{fnx\_06\_019}&\multicolumn{2}{c}{fnx0579x--1}\\
 
\hline\Tstrut
\ion{Al}{ I} & 3961.52 & 0.01 & $-0.323$ & (162.4) & $ 3.56$ & -- & -- \\
\hline\Tstrut
\ion{Ba}{II} & 4934.076 & 0.0 & $-0.150$ & (192.8) & $-0.39$ & ( 93.2) & $-1.35$ \\
\ion{Ba}{II} & 5853.668 & 0.6 & $-1.000$ & ( 70.4) & $-0.65$ & ( 35.9) & $-1.17$ \\
\ion{Ba}{II} & 6141.713 & 0.7 & $-0.076$ & (126.8) & $-0.45$ & ( 83.9) & $-1.23$ \\
\ion{Ba}{II} & 6496.897 & 0.6 & $-0.377$ & (136.7) & $-0.22$ & ( 81.5) & $-1.10$ \\
\hline\Tstrut
\ion{ C}{ I} & 4325.0 & $--$ & $--$ & -- & $ 5.05$ & -- & $ 5.46$ \\
\hline\Tstrut
\ion{Ca}{ I} & 5581.965 & 2.52 & $-0.555$ & -- & -- & 25.2 $\pm$  3.6 & $ 3.86$ \\
\ion{Ca}{ I} & 5588.749 & 2.53 & $ 0.358$ & -- & -- & 67.7 $\pm$  6.1 & $ 3.73$ \\
\ion{Ca}{ I} & 5590.114 & 2.52 & $-0.571$ & -- & -- & 31.9 $\pm$  3.8 & $ 4.02$ \\
\ion{Ca}{ I} & 5601.277 & 2.53 & $-0.523$ & -- & -- & 27.8 $\pm$  4.3 & $ 3.89$ \\
\ion{Ca}{ I} & 5857.451 & 2.93 & $ 0.240$ & 33.6 $\pm$  4.2 & $ 3.80$ & 41.0 $\pm$  5.0 & $ 3.90$ \\
\ion{Ca}{ I} & 6102.723 & 1.88 & $-0.793$ & 55.4 $\pm$  5.6 & $ 3.78$ & 71.1 $\pm$  6.3 & $ 3.97$ \\
\ion{Ca}{ I} & 6122.217 & 1.89 & $-0.316$ & 88.3 $\pm$  7.4 & $ 3.81$ & 98.8 $\pm$  9.1 & $ 3.93$ \\
\ion{Ca}{ I} & 6162.173 & 1.9 & $-0.090$ & 105.6 $\pm$  7.5 & $ 3.87$ & -- & -- \\
\ion{Ca}{ I} & 6169.042 & 2.52 & $-0.797$ & -- & -- & 20.6 $\pm$  3.9 & $ 3.95$ \\
\ion{Ca}{ I} & 6169.563 & 2.53 & $-0.478$ & 27.4 $\pm$  3.6 & $ 3.84$ & 31.4 $\pm$  3.4 & $ 3.87$ \\
\ion{Ca}{ I} & 6439.075 & 2.53 & $ 0.390$ & 79.3 $\pm$  8.4 & $ 3.82$ & 88.5 $\pm$  7.6 & $ 3.93$ \\
\ion{Ca}{ I} & 6717.681 & 2.71 & $-0.524$ & -- & -- & 21.6 $\pm$  4.4 & $ 3.91$ \\
\hline\Tstrut
\ion{Co}{ I} & 4121.318 & 0.92 & $-0.320$ & 105.6 $\pm$  6.3 & $ 2.12$ & (130.5) & $ 2.07$ \\
\hline\Tstrut
\ion{Cr}{ I} & 5296.691 & 0.98 & $-1.360$ & 36.9 $\pm$  4.7 & $ 2.51$ & 48.9 $\pm$  5.6 & $ 2.62$ \\
\ion{Cr}{ I} & 5298.271 & 0.98 & $-1.140$ & -- & -- & 63.6 $\pm$  6.3 & $ 2.63$ \\
\ion{Cr}{ I} & 5345.796 & 1.0 & $-0.896$ & 56.2 $\pm$  6.0 & $ 2.39$ & 72.5 $\pm$  5.4 & $ 2.55$ \\
\ion{Cr}{ I} & 5348.314 & 1.0 & $-1.210$ & 34.4 $\pm$  4.8 & $ 2.34$ & 59.6 $\pm$  4.5 & $ 2.66$ \\
\ion{Cr}{ I} & 5409.784 & 1.03 & $-0.670$ & 64.2 $\pm$  5.7 & $ 2.31$ & 90.3 $\pm$  8.4 & $ 2.64$ \\
\hline\Tstrut
\ion{Cu}{ I} & 5105.537 & 1.39 & $-1.542$ & -- & $<0.41$ & ( 20.7) & $ 0.82$ \\
\hline\Tstrut
\ion{Dy}{II} & 3944.68 & 0.0 & $ 0.000$ & -- & $-0.74$ & -- & -- \\
\ion{Dy}{II} & 4103.31 & 0.0 & $ 0.000$ & -- & $-0.74$ & -- & -- \\
\ion{Dy}{II} & 4449.7  & 0.0 & $ 0.000$ & -- & $-0.74$ & -- & -- \\
\hline\Tstrut
\ion{Eu}{II} & 4129.708 & 0.0 & $ 0.220$ & ( 84.6) & $-1.56$ & -- & -- \\
\ion{Eu}{II} & 4205.042 & 0.0 & $ 0.210$ & (143.7) & $-1.65$ & -- & -- \\
\hline\Tstrut
\ion{Fe}{ I} & 4859.741 & 2.88 & $-0.764$ & 73.9 $\pm$  6.3 & $ 4.50$ & 79.6 $\pm$  7.8 & $ 4.58$ \\
\ion{Fe}{ I} & 4871.318 & 2.87 & $-0.363$ & 100.4 $\pm$  9.8 & $ 4.60$ & -- & -- \\
\ion{Fe}{ I} & 4872.138 & 2.88 & $-0.567$ & 87.0 $\pm$  8.3 & $ 4.56$ & 106.4 $\pm$  9.9 & $ 4.93$ \\
\ion{Fe}{ I} & 4890.755 & 2.88 & $-0.394$ & 101.7 $\pm$  8.3 & $ 4.66$ & 104.8 $\pm$  9.7 & $ 4.71$ \\
\ion{Fe}{ I} & 4891.492 & 2.85 & $-0.112$ & 104.7 $\pm$ 11.2 & $ 4.41$ & -- & -- \\
\ion{Fe}{ I} & 4903.31 & 2.88 & $-0.926$ & 78.8 $\pm$  8.4 & $ 4.75$ & 85.1 $\pm$  8.8 & $ 4.85$ \\
\ion{Fe}{ I} & 4924.77 & 2.28 & $-2.241$ & 53.1 $\pm$  6.1 & $ 4.77$ & 67.3 $\pm$  8.2 & $ 4.97$ \\
\ion{Fe}{ I} & 4938.814 & 2.88 & $-1.077$ & 58.4 $\pm$  7.7 & $ 4.51$ & 73.9 $\pm$  8.5 & $ 4.76$ \\
\ion{Fe}{ I} & 4966.088 & 3.33 & $-0.871$ & 45.0 $\pm$  5.5 & $ 4.68$ & 58.5 $\pm$  8.2 & $ 4.89$ \\
\ion{Fe}{ I} & 5001.863 & 3.88 & $ 0.010$ & 41.9 $\pm$  4.7 & $ 4.47$ & 49.7 $\pm$  5.2 & $ 4.58$ \\
\ion{Fe}{ I} & 5006.119 & 2.83 & $-0.638$ & 90.2 $\pm$ 10.5 & $ 4.59$ & -- & -- \\
\ion{Fe}{ I} & 5014.942 & 3.94 & $-0.303$ & 38.9 $\pm$  4.5 & $ 4.80$ & 35.8 $\pm$  7.0 & $ 4.71$ \\
\ion{Fe}{ I} & 5044.211 & 2.85 & $-2.038$ & -- & -- & 32.1 $\pm$  5.9 & $ 4.90$ \\
\ion{Fe}{ I} & 5049.82 & 2.28 & $-1.355$ & 93.8 $\pm$  9.4 & $ 4.59$ & 101.9 $\pm$ 11.0 & $ 4.72$ \\
\ion{Fe}{ I} & 5068.766 & 2.94 & $-1.042$ & 46.3 $\pm$  4.7 & $ 4.33$ & 73.7 $\pm$  7.9 & $ 4.78$ \\
\ion{Fe}{ I} & 5074.748 & 4.22 & $-0.200$ & -- & -- & 24.7 $\pm$  3.5 & $ 4.72$ \\
\ion{Fe}{ I} & 5079.223 & 2.2 & $-2.067$ & 73.5 $\pm$  6.5 & $ 4.81$ & 86.1 $\pm$  7.8 & $ 5.00$ \\
\ion{Fe}{ I} & 5131.468 & 2.22 & $-2.515$ & 37.2 $\pm$  4.7 & $ 4.67$ & 50.7 $\pm$  5.0 & $ 4.84$ \\
\ion{Fe}{ I} & 5141.739 & 2.42 & $-1.964$ & 38.3 $\pm$  4.8 & $ 4.41$ & 48.1 $\pm$  4.7 & $ 4.52$ \\
\ion{Fe}{ I} & 5145.094 & 2.2 & $-2.876$ & -- & -- & 20.7 $\pm$  3.9 & $ 4.57$ \\
\ion{Fe}{ I} & 5162.272 & 4.18 & $ 0.020$ & 30.4 $\pm$  4.0 & $ 4.60$ & 40.9 $\pm$  4.5 & $ 4.78$ \\
\ion{Fe}{ I} & 5191.455 & 3.04 & $-0.551$ & 78.1 $\pm$  8.0 & $ 4.52$ & 89.6 $\pm$  8.5 & $ 4.71$ \\
\ion{Fe}{ I} & 5192.344 & 3.0 & $-0.421$ & 86.1 $\pm$  7.9 & $ 4.48$ & 102.5 $\pm$  9.0 & $ 4.77$ \\
\ion{Fe}{ I} & 5198.711 & 2.22 & $-2.135$ & 57.4 $\pm$  6.1 & $ 4.62$ & 70.3 $\pm$  5.8 & $ 4.78$ \\
\ion{Fe}{ I} & 5202.336 & 2.18 & $-1.838$ & 83.4 $\pm$  6.7 & $ 4.70$ & -- & -- \\
\ion{Fe}{ I} & 5215.18 & 3.27 & $-0.871$ & 44.0 $\pm$  4.1 & $ 4.54$ & -- & -- \\
\ion{Fe}{ I} & 5216.274 & 1.61 & $-2.150$ & 99.9 $\pm$ 11.6 & $ 4.50$ & -- & -- \\
\ion{Fe}{ I} & 5217.389 & 3.21 & $-1.070$ & 33.4 $\pm$  4.8 & $ 4.47$ & 56.4 $\pm$  5.4 & $ 4.84$ \\
\ion{Fe}{ I} & 5242.491 & 3.63 & $-0.967$ & -- & -- & 27.8 $\pm$  6.1 & $ 4.77$ \\
\ion{Fe}{ I} & 5266.555 & 3.0 & $-0.386$ & -- & -- & 101.9 $\pm$ 10.5 & $ 4.70$ \\
\ion{Fe}{ I} & 5281.79 & 3.04 & $-0.834$ & 50.8 $\pm$  8.7 & $ 4.31$ & 69.0 $\pm$  7.0 & $ 4.58$ \\
\ion{Fe}{ I} & 5283.621 & 3.24 & $-0.432$ & 69.2 $\pm$  8.0 & $ 4.50$ & 87.4 $\pm$  9.4 & $ 4.81$ \\
\ion{Fe}{ I} & 5302.3 & 3.28 & $-0.720$ & 45.9 $\pm$  5.4 & $ 4.44$ & 60.3 $\pm$  7.3 & $ 4.65$ \\
\ion{Fe}{ I} & 5307.361 & 1.61 & $-2.987$ & 65.1 $\pm$  5.9 & $ 4.72$ & 77.3 $\pm$  6.1 & $ 4.86$ \\
\ion{Fe}{ I} & 5322.041 & 2.28 & $-2.803$ & 20.3 $\pm$  3.9 & $ 4.65$ & 26.1 $\pm$  4.4 & $ 4.72$ \\
\ion{Fe}{ I} & 5324.179 & 3.21 & $-0.103$ & 87.6 $\pm$  7.2 & $ 4.46$ & 97.5 $\pm$  9.5 & $ 4.62$ \\
\ion{Fe}{ I} & 5332.899 & 1.56 & $-2.777$ & 78.8 $\pm$  8.9 & $ 4.66$ & 92.5 $\pm$  7.4 & $ 4.84$ \\
\ion{Fe}{ I} & 5339.929 & 3.27 & $-0.647$ & 53.0 $\pm$  5.9 & $ 4.46$ & 70.4 $\pm$  7.8 & $ 4.73$ \\
\ion{Fe}{ I} & 5364.871 & 4.45 & $ 0.228$ & -- & -- & 32.1 $\pm$  3.3 & $ 4.73$ \\

\end{tabular}}
\label{Tab:lines}
\end{table}

\begin{table}[htp]
\resizebox{\linewidth}{!}{
\begin{tabular}{lllr|rrrr}

\hline
\hline\Tstrut
El. & $\lambda$ & $\chi_{ex} $ & log($gf$) & EW [m\AA] & log$\epsilon$(X) & EW [m\AA] & log$\epsilon$(X) \\
 & [\AA] & [eV] &  &\multicolumn{2}{c}{fnx\_06\_019}&\multicolumn{2}{c}{fnx0579x--1}\\

\hline\Tstrut
\ion{Fe}{ I} & 5365.399 & 3.57 & $-1.020$ & -- & -- & 21.9 $\pm$  4.5 & $ 4.59$ \\
\ion{Fe}{ I} & 5367.466 & 4.41 & $ 0.443$ & -- & -- & 41.8 $\pm$  6.1 & $ 4.67$ \\
\ion{Fe}{ I} & 5369.961 & 4.37 & $ 0.536$ & 37.6 $\pm$  5.0 & $ 4.46$ & 43.9 $\pm$  4.3 & $ 4.55$ \\
\ion{Fe}{ I} & 5383.369 & 4.31 & $ 0.645$ & 41.3 $\pm$  4.0 & $ 4.35$ & 58.3 $\pm$  7.8 & $ 4.63$ \\
\ion{Fe}{ I} & 5393.167 & 3.24 & $-0.715$ & 51.1 $\pm$  5.3 & $ 4.46$ & 67.1 $\pm$  5.8 & $ 4.69$ \\
\ion{Fe}{ I} & 5410.91 & 4.47 & $ 0.398$ & 27.2 $\pm$  4.2 & $ 4.52$ & 33.8 $\pm$  3.9 & $ 4.63$ \\
\ion{Fe}{ I} & 5415.199 & 4.39 & $ 0.642$ & 45.1 $\pm$  5.3 & $ 4.51$ & 54.2 $\pm$  5.6 & $ 4.65$ \\
\ion{Fe}{ I} & 5424.068 & 4.32 & $ 0.520$ & 45.1 $\pm$  5.1 & $ 4.55$ & 56.0 $\pm$  6.7 & $ 4.72$ \\
\ion{Fe}{ I} & 5445.042 & 4.39 & $-0.020$ & 20.0 $\pm$  2.7 & $ 4.64$ & 30.3 $\pm$  3.2 & $ 4.86$ \\
\ion{Fe}{ I} & 5569.618 & 3.42 & $-0.486$ & 52.9 $\pm$  4.8 & $ 4.47$ & -- & -- \\
\ion{Fe}{ I} & 5572.842 & 3.4 & $-0.275$ & 68.0 $\pm$  7.0 & $ 4.49$ & 80.7 $\pm$  7.6 & $ 4.68$ \\
\ion{Fe}{ I} & 5586.755 & 3.37 & $-0.120$ & 80.4 $\pm$  6.9 & $ 4.51$ & 96.9 $\pm$  9.5 & $ 4.78$ \\
\ion{Fe}{ I} & 5615.644 & 3.33 & $ 0.050$ & 101.2 $\pm$  8.8 & $ 4.65$ & 106.0 $\pm$  8.9 & $ 4.72$ \\
\ion{Fe}{ I} & 5701.544 & 2.56 & $-2.216$ & 29.0 $\pm$  3.9 & $ 4.61$ & 49.3 $\pm$  4.7 & $ 4.91$ \\
\ion{Fe}{ I} & 6065.482 & 2.61 & $-1.530$ & 67.9 $\pm$  5.9 & $ 4.60$ & 84.5 $\pm$  7.6 & $ 4.81$ \\
\ion{Fe}{ I} & 6136.615 & 2.45 & $-1.400$ & 89.7 $\pm$  7.9 & $ 4.58$ & -- & -- \\
\ion{Fe}{ I} & 6137.691 & 2.59 & $-1.403$ & 72.3 $\pm$  7.0 & $ 4.50$ & 95.5 $\pm$  8.6 & $ 4.82$ \\
\ion{Fe}{ I} & 6151.617 & 2.18 & $-3.299$ & -- & -- & 20.7 $\pm$  3.5 & $ 4.87$ \\
\ion{Fe}{ I} & 6173.334 & 2.22 & $-2.880$ & 21.8 $\pm$  2.1 & $ 4.62$ & 32.1 $\pm$  4.0 & $ 4.76$ \\
\ion{Fe}{ I} & 6191.558 & 2.43 & $-1.417$ & 84.1 $\pm$  6.8 & $ 4.47$ & 98.8 $\pm$  9.0 & $ 4.66$ \\
\ion{Fe}{ I} & 6200.312 & 2.61 & $-2.437$ & -- & -- & 26.7 $\pm$  3.7 & $ 4.74$ \\
\ion{Fe}{ I} & 6213.429 & 2.22 & $-2.482$ & 43.5 $\pm$  5.2 & $ 4.62$ & 56.0 $\pm$  5.6 & $ 4.75$ \\
\ion{Fe}{ I} & 6219.28 & 2.2 & $-2.433$ & 52.0 $\pm$  4.9 & $ 4.67$ & 70.2 $\pm$  6.5 & $ 4.88$ \\
\ion{Fe}{ I} & 6230.722 & 2.56 & $-1.281$ & 98.2 $\pm$  7.6 & $ 4.73$ & -- & -- \\
\ion{Fe}{ I} & 6240.646 & 2.22 & $-3.233$ & -- & -- & 25.3 $\pm$  3.9 & $ 4.97$ \\
\ion{Fe}{ I} & 6246.318 & 3.6 & $-0.733$ & 31.3 $\pm$  3.7 & $ 4.53$ & 46.2 $\pm$  5.3 & $ 4.75$ \\
\ion{Fe}{ I} & 6252.555 & 2.4 & $-1.687$ & 77.1 $\pm$  6.2 & $ 4.58$ & 91.0 $\pm$  9.3 & $ 4.75$ \\
\ion{Fe}{ I} & 6265.132 & 2.18 & $-2.550$ & 43.1 $\pm$  4.7 & $ 4.62$ & 62.3 $\pm$  5.5 & $ 4.84$ \\
\ion{Fe}{ I} & 6297.792 & 2.22 & $-2.740$ & -- & -- & 49.4 $\pm$  8.6 & $ 4.90$ \\
\ion{Fe}{ I} & 6301.5 & 3.65 & $-0.718$ & -- & -- & 59.5 $\pm$  7.4 & $ 5.01$ \\
\ion{Fe}{ I} & 6302.494 & 3.69 & $-0.973$ & -- & -- & 22.1 $\pm$  2.2 & $ 4.63$ \\
\ion{Fe}{ I} & 6322.685 & 2.59 & $-2.426$ & 26.7 $\pm$  3.3 & $ 4.76$ & 37.8 $\pm$  4.9 & $ 4.90$ \\
\ion{Fe}{ I} & 6335.33 & 2.2 & $-2.177$ & 66.0 $\pm$  6.7 & $ 4.61$ & 83.0 $\pm$  7.0 & $ 4.80$ \\
\ion{Fe}{ I} & 6344.148 & 2.43 & $-2.923$ & -- & -- & 23.0 $\pm$  3.2 & $ 4.89$ \\
\ion{Fe}{ I} & 6355.028 & 2.85 & $-2.350$ & -- & -- & 22.1 $\pm$  3.7 & $ 4.86$ \\
\ion{Fe}{ I} & 6393.601 & 2.43 & $-1.432$ & 84.5 $\pm$  6.2 & $ 4.46$ & 100.0 $\pm$  9.5 & $ 4.65$ \\
\ion{Fe}{ I} & 6400.0 & 3.6 & $-0.290$ & 53.7 $\pm$ 10.8 & $ 4.46$ & 73.6 $\pm$ 13.8 & $ 4.73$ \\
\ion{Fe}{ I} & 6408.018 & 3.69 & $-1.018$ & -- & -- & 28.1 $\pm$  3.7 & $ 4.80$ \\
\ion{Fe}{ I} & 6411.648 & 3.65 & $-0.595$ & 32.7 $\pm$  3.6 & $ 4.47$ & 48.8 $\pm$  6.1 & $ 4.71$ \\
\ion{Fe}{ I} & 6421.35 & 2.28 & $-2.027$ & 75.6 $\pm$  6.8 & $ 4.70$ & 95.9 $\pm$  7.5 & $ 4.96$ \\
\ion{Fe}{ I} & 6430.845 & 2.18 & $-2.006$ & 90.5 $\pm$  6.7 & $ 4.75$ & 104.1 $\pm$  7.7 & $ 4.90$ \\
\ion{Fe}{ I} & 6592.913 & 2.73 & $-1.473$ & 62.2 $\pm$  5.2 & $ 4.56$ & 79.9 $\pm$  7.0 & $ 4.77$ \\
\ion{Fe}{ I} & 6593.87 & 2.43 & $-2.422$ & 33.1 $\pm$  3.7 & $ 4.64$ & 45.3 $\pm$  4.2 & $ 4.78$ \\
\ion{Fe}{ I} & 6609.11 & 2.56 & $-2.692$ & -- & -- & 23.4 $\pm$  2.4 & $ 4.82$ \\
\ion{Fe}{ I} & 6677.985 & 2.69 & $-1.418$ & 80.3 $\pm$  6.5 & $ 4.71$ & 104.0 $\pm$  9.9 & $ 5.03$ \\
\ion{Fe}{ I} & 6750.151 & 2.42 & $-2.621$ & 28.6 $\pm$  3.7 & $ 4.73$ & 44.4 $\pm$  5.1 & $ 4.94$ \\
\hline\Tstrut
\ion{Fe}{II} & 5197.567 & 3.23 & $-2.100$ & 45.5 $\pm$  4.9 & $ 4.68$ & 53.5 $\pm$  7.8 & $ 4.88$ \\
\ion{Fe}{II} & 5234.623 & 3.22 & $-2.230$ & -- & -- & 54.6 $\pm$  7.8 & $ 5.02$ \\
\ion{Fe}{II} & 5284.103 & 2.89 & $-2.990$ & 23.3 $\pm$  4.0 & $ 4.66$ & 31.0 $\pm$  5.6 & $ 4.86$ \\
\ion{Fe}{II} & 6516.077 & 2.89 & $-3.320$ & -- & -- & 28.5 $\pm$  4.4 & $ 5.06$ \\
\hline\Tstrut
\ion{La}{II} & 4920.98 & 0.13 & $-0.580$ & ( 37.6) & $-1.28$ & -- & -- \\
\ion{La}{II} & 3995.75 & 0.0 & $ 0.000$ & -- & $-1.28$ & -- & -- \\
\ion{La}{II} & 4077.34 & 0.0 & $ 0.000$ & -- & $-1.28$ & -- & -- \\
\ion{La}{II} & 4123.23 & 0.0 & $ 0.000$ & -- & $-1.28$ & -- & -- \\
\hline\Tstrut
\ion{Mg}{ I} & 5172.684 & 2.71 & $-0.450$ & (235.1) & $ 5.17$ & (134.2) & $ 5.17$ \\
\ion{Mg}{ I} & 5183.604 & 2.72 & $-0.239$ & (180.3) & $ 5.27$ & (163.9) & $ 5.20$ \\
\ion{Mg}{ I} & 5528.405 & 4.35 & $-0.498$ & 90.4 $\pm$  9.3 & $ 5.26$ & ( 55.9) & $ 5.25$ \\
\hline\Tstrut
\ion{Mn}{ I} & 4030.75 & 0.0 & $-0.494$ & (162.2) & $ 2.06$ & -- & -- \\
\ion{Mn}{ I} & 4033.06 & 0.0 & $-0.644$ & (171.8) & $ 2.10$ & (191.7) & $ 2.27$ \\
\ion{Mn}{ I} & 4034.48 & 0.0 & $-0.842$ & (125.0) & $ 2.10$ & (103.0) & $ 2.27$ \\
\ion{Mn}{ I} & 4823.52 & 2.32 & $ 0.121$ & -- & -- & 45.3 $\pm$  5.9 & $ 2.17$ \\
\hline\Tstrut
\ion{Na}{ I} & 5889.951 & 0.0 & $ 0.108$ & (208.8) & $ 3.52$ & (207.0) & $ 3.48$ \\
\ion{Na}{ I} & 5895.924 & 0.0 & $-0.194$ & (186.0) & $ 3.57$ & (186.3) & $ 3.53$ \\
\hline\Tstrut
\ion{Nd}{II} & 4825.48 & 0.18 & $-0.420$ & ( 41.4) & $-0.89$ & -- & -- \\
\ion{Nd}{II} & 5319.81 & 0.55 & $-0.140$ & ( 28.4) & $-0.89$ & -- & -- \\
\ion{Nd}{II} & 4061.08 & 0.0 & $ 0.000$ & -- & $-0.89$ & -- & -- \\
\hline\Tstrut
\ion{Ni}{ I} & 5081.11 & 3.85 & $ 0.300$ & -- & -- & 20.9 $\pm$  4.2 & $ 3.39$ \\
\ion{Ni}{ I} & 5476.904 & 1.83 & $-0.780$ & 101.0 $\pm$  7.6 & $ 3.20$ & 107.0 $\pm$ 10.1 & $ 3.29$ \\
\ion{Ni}{ I} & 6643.63 & 1.68 & $-2.220$ & 40.2 $\pm$  5.0 & $ 3.29$ & 48.7 $\pm$  5.7 & $ 3.36$ \\
\ion{Ni}{ I} & 6767.772 & 1.83 & $-2.140$ & 39.8 $\pm$  3.8 & $ 3.40$ & 44.2 $\pm$  4.4 & $ 3.41$ \\
\hline\Tstrut
\ion{ O}{ I} & 6300.304 & 0.0 & $-9.750$ & ( 31.0) & $ 6.67$ & ( 31.0) & $ 6.99$ \\
\hline\Tstrut
\ion{Pr}{II} & 4222.934 & 0.05 & $ 0.271$ & 60.6 $\pm$  7.3 & $-1.14$ & -- & -- \\
\hline\Tstrut
\ion{Sc}{II} & 4246.822 & 0.31 & $ 0.242$ & -- & -- & (167.7) & $ 0.73$ \\
\ion{Sc}{II} & 4400.389 & 0.61 & $-0.536$ & ( 94.5) & $ 0.46$ & -- & -- \\
\ion{Sc}{II} & 4415.557 & 0.6 & $-0.668$ & ( 88.4) & $ 0.35$ & (104.4) & $ 0.90$ \\
\ion{Sc}{II} & 5031.021 & 1.36 & $-0.400$ & ( 49.6) & $ 0.30$ & ( 51.4) & $ 0.35$ \\
\ion{Sc}{II} & 5526.79 & 1.77 & $ 0.024$ & ( 49.4) & $ 0.39$ & ( 58.6) & $ 0.57$ \\
\ion{Sc}{II} & 5657.896 & 1.51 & $-0.603$ & ( 42.9) & $ 0.51$ & ( 51.7) & $ 0.67$ \\
\ion{Sc}{II} & 6604.601 & 1.36 & $-1.309$ & ( 17.4) & $ 0.40$ & -- & -- \\
\hline\Tstrut
\ion{Si}{ I} & 4102.936 & 1.91 & $-3.140$ & (122.7) & $ 5.16$ & ( 78.8) & $ 4.76$ \\
\hline\Tstrut
\ion{Sm}{II} & 4424.34 & 0.49 & $ 0.140$ & 42.5 $\pm$  6.8 & $-1.12$ & -- & -- \\
 
\end{tabular}}
\end{table}

\begin{table}[htp]
\setcounter{table}{0}
\centering
\caption{continued.}
\resizebox{\linewidth}{!}{
%\resizebox{0.7\linewidth}{!}{
\begin{tabular}{lllr|rrrr}

\hline
\hline\Tstrut
El. & $\lambda$ & $\chi_{ex} $ & log($gf$) & EW [m\AA] & log$\epsilon$(X) & EW [m\AA] & log$\epsilon$(X) \\
 & [\AA] & [eV] &  &\multicolumn{2}{c}{fnx\_06\_019}&\multicolumn{2}{c}{fnx0579x--1}\\

\hline\Tstrut
\ion{Sr}{II} & 4215.519 & 0.0 & $-0.145$ & (170.7) & $-0.67$ & (126.4) & $-0.52$ \\
\hline\Tstrut
\ion{Ti}{ I} & 4840.874 & 0.9 & $-0.430$ & -- & -- & 30.4 $\pm$  4.7 & $ 2.09$ \\
\ion{Ti}{ I} & 4981.73 & 0.85 & $ 0.570$ & 85.8 $\pm$  8.2 & $ 2.01$ & 93.8 $\pm$  9.1 & $ 2.06$ \\
\ion{Ti}{ I} & 4991.066 & 0.84 & $ 0.450$ & 82.1 $\pm$  9.6 & $ 2.05$ & 80.0 $\pm$ 10.1 & $ 1.92$ \\
\ion{Ti}{ I} & 4999.503 & 0.83 & $ 0.320$ & 68.9 $\pm$  7.6 & $ 1.95$ & 72.0 $\pm$  7.9 & $ 1.90$ \\
\ion{Ti}{ I} & 5016.161 & 0.85 & $-0.480$ & 25.3 $\pm$  3.9 & $ 2.04$ & -- & -- \\
\ion{Ti}{ I} & 5039.958 & 0.02 & $-1.080$ & 54.5 $\pm$  4.8 & $ 1.94$ & 64.6 $\pm$  7.0 & $ 1.95$ \\
\ion{Ti}{ I} & 5064.653 & 0.05 & $-0.940$ & 61.5 $\pm$  6.5 & $ 1.94$ & 79.9 $\pm$  8.2 & $ 2.07$ \\
\ion{Ti}{ I} & 5173.743 & 0.0 & $-1.060$ & 64.8 $\pm$  5.9 & $ 2.01$ & 69.9 $\pm$  8.1 & $ 1.94$ \\
\ion{Ti}{ I} & 5192.969 & 0.02 & $-0.950$ & 71.3 $\pm$  6.3 & $ 2.02$ & 79.0 $\pm$ 10.0 & $ 1.99$ \\
\ion{Ti}{ I} & 5210.384 & 0.05 & $-0.820$ & 67.8 $\pm$  6.5 & $ 1.88$ & 72.7 $\pm$ 11.4 & $ 1.81$ \\
\hline\Tstrut
\ion{Ti}{II} & 4798.531 & 1.08 & $-2.660$ & 43.6 $\pm$  7.3 & $ 2.49$ & -- & -- \\
\ion{Ti}{II} & 5129.156 & 1.89 & $-1.340$ & 53.4 $\pm$  6.3 & $ 2.39$ & 64.4 $\pm$  7.2 & $ 2.63$ \\
\ion{Ti}{II} & 5154.068 & 1.57 & $-1.750$ & 61.5 $\pm$  7.0 & $ 2.50$ & 62.2 $\pm$  6.3 & $ 2.55$ \\
\ion{Ti}{II} & 5185.902 & 1.89 & $-1.410$ & 49.2 $\pm$  4.4 & $ 2.38$ & 54.7 $\pm$  6.1 & $ 2.51$ \\
\ion{Ti}{II} & 5188.687 & 1.58 & $-1.050$ & -- & -- & 109.4 $\pm$ 12.8 & $ 2.81$ \\
\ion{Ti}{II} & 5226.539 & 1.57 & $-1.260$ & 94.0 $\pm$  7.9 & $ 2.61$ & 93.6 $\pm$  8.2 & $ 2.66$ \\
\ion{Ti}{II} & 5336.786 & 1.58 & $-1.600$ & -- & -- & 63.7 $\pm$ 10.6 & $ 2.42$ \\
\ion{Ti}{II} & 5381.021 & 1.57 & $-1.970$ & 53.2 $\pm$  6.2 & $ 2.55$ & 45.1 $\pm$  6.5 & $ 2.43$ \\
\ion{Ti}{II} & 5418.768 & 1.58 & $-2.130$ & 34.5 $\pm$  5.2 & $ 2.40$ & 39.8 $\pm$  4.5 & $ 2.52$ \\
\hline\Tstrut
\ion{ Y}{II} & 4883.682 & 1.08 & $ 0.070$ & ( 36.5) & $-0.92$ & ( 23.1) & $-1.18$ \\
\ion{ Y}{II} & 4900.119 & 1.03 & $-0.090$ & ( 38.0) & $-0.78$ & -- & -- \\
\ion{ Y}{II} & 5087.419 & 1.08 & $-0.170$ & ( 26.2) & $-0.92$ & ( 27.0) & $-0.89$ \\
\ion{ Y}{II} & 5200.41 & 0.99 & $-0.570$ & ( 18.0) & $-0.86$ & -- & -- \\
\ion{ Y}{II} & 5205.722 & 1.03 & $-0.340$ & ( 31.0) & $-0.80$ & -- & -- \\
\hline\Tstrut
\ion{Zn}{ I} & 4810.528 & 4.08 & $-0.137$ & ( 24.4) & $ 1.93$ & -- & $<1.91$ \\
\hline\Tstrut
\ion{Zr}{II} & 4161.2 & 0.71 & $-0.590$ & ( 42.9) & $-0.12$ & -- & -- \\

\end{tabular}}
\end{table}

%%%%%%%%%%%%%%%%%%%%%%%%%%%%%%%%%%%%%%%%%%%%%%%%%%%%%%%%%%%%%%%%%%%%%%%%%%%%%%%%%%%%%%%%%%%%%
\begin{table}[htp]
\vspace{2cm}
\centering
\caption{Computed NLTE corrections for Na.}
\resizebox{\linewidth}{!}{
\begin{tabular}{c|cccccc}
\hline
\hline\Tstrut
ID & fnx\_06\_019 & fnx0579x-1 & car1\_t174 & car1\_t194 & car1\_t200 & LG04c\_0008 \\
\hline\Tstrut
     & \multicolumn{6}{c}{5889.951~\AA} \\
\hline\Tstrut
log$\epsilon$(Na)$_{LTE}$  & 3.52    & 3.48    & 3.80    & 4.16    & 4.52    & 3.78    \\
log$\epsilon$(Na)$_{NLTE}$ & 3.37    & 3.36    & 3.42    & 3.96    & 4.25    & 3.43    \\
$\Delta$(NLTE-LTE)         & $-0.15$ & $-0.12$ & $-0.38$ & $-0.20$ & $-0.27$ & $-0.35$ \\ 
$[Na/Fe]_{NLTE}$           & $+0.04$ & $-0.16$ & $+0.18$ & $+0.26$ & $+0.95$ & $+0.23$ \\ 
\hline\Tstrut
     & \multicolumn{6}{c}{5895.924~\AA}  \\
\hline\Tstrut
log$\epsilon$(Na)$_{LTE}$  & 3.57    & 3.53    & $--$    & 4.09    & 4.35    & 4.00    \\
log$\epsilon$(Na)$_{NLTE}$ & 3.38    & 3.37    & $--$    & 3.79    & 3.95    & 3.61    \\
$\Delta$(NLTE-LTE)         & $-0.19$ & $-0.16$ & $--$    & $-0.30$ & $-0.40$ & $-0.39$ \\
$[Na/Fe]_{NLTE}$           & $+0.05$ & $-0.15$ & $--$    & $+0.09$ & $+0.65$ & $+0.41$ \\
\hline
\end{tabular}}
\label{Tab:NLTE_corrections}
\end{table}
%%%%%%%%%%%%%%%%%%%%%%%%%%%%%%%%%%%%%%%%%%%%%%%%%%%%%%%%%%%%%%%%%%%%%%%%%%%%%%%%%%%%%%%%%%%%%

%%%%%%%%%%%%%%%%%%%%%%%%%%%%%%%%%%%%%%%%%%%%%%%%%%%%%%%%%%%%%%%%%%%%%%%%%%%%%%%%%%%%%%%%%%%%%
% XSHOOTER
%%%%%%%%%%%%%%%%%%%%%%%%%%%%%%%%%%%%%%%%%%%%%%%%%%%%%%%%%%%%%%%%%%%%%%%%%%%%%%%%%%%%%%%%%%%%%

\clearpage
\onecolumn

\begin{table*}[ht]
\centering
\caption{Lines measured in the Carina XSHOOTER spectra. Line parameters, observed EWs, and elemental abundances are provided. The EWs in brackets are given only as an indication; the quoted abundances are derived through spectral synthesis for these lines.}
\resizebox{0.79\linewidth}{!}{
\begin{tabular}{lllr|rrrrrrrr}

\hline\Tstrut
El. & $\lambda$ & $\chi_{ex} $ & log($gf$) & EW [m\AA] & log$\epsilon$(X) & EW [m\AA] & log$\epsilon$(X) & EW [m\AA] & log$\epsilon$(X) & EW [m\AA] & log$\epsilon$(X) \\
 & [\AA] & [eV] &  &\multicolumn{2}{c}{car1\_t174}&\multicolumn{2}{c}{car1\_t194}&\multicolumn{2}{c}{car1\_t200}&\multicolumn{2}{c}{LG04c\_0008}\\
\hline

\Tstrut
\ion{Ba}{II} & 4554.029 & 0.0 & $ 0.170$ & ( 79.2) & $-1.73$ & (165.8) & $ 0.14$ & (139.2) & $-0.32$ & ( 84.7) & $-1.80$ \\
\ion{Ba}{II} & 4934.076 & 0.0 & $-0.150$ & ( 38.4) & $-2.25$ & (164.7) & $-0.08$ & (115.0) & $-0.87$ & ( 51.9) & $-2.23$ \\
\ion{ C}{ I} & 4300.0   & -- & -- & -- & $ 4.59$ & -- & $ 4.90$ & -- & $ 5.06$ & -- & $ 4.22$ \\
\ion{Ca}{ I} & 4226.728 & 0.0 & $ 0.244$ & 239.1 $\pm$  13.6 & $ 3.64$ & -- & -- & -- & -- & -- & -- \\
\ion{Ca}{ I} & 6122.217 & 1.89 & $-0.316$ & -- & -- & 78.9 $\pm$  15.2 & $ 3.96$ & -- & -- & -- & -- \\
\ion{Ca}{ I} & 6439.075 & 2.53 & $ 0.390$ & -- & -- & 92.4 $\pm$  16.5 & $ 4.30$ & -- & -- & 48.1 $\pm$   8.5 & $ 3.55$ \\
\ion{Co}{ I} & 4118.773 & 1.05 & $-0.490$ & -- & -- & -- & -- & -- & -- & (102.1) & $ 1.81$ \\
\ion{Co}{ I} & 4121.318 & 0.92 & $-0.320$ & -- & -- & -- & -- & -- & -- & (107.7) & $ 1.75$ \\
\ion{Cr}{ I} & 5206     & 0.94 & $ 0.020$ & -- & $ 2.07$ & -- & $ 2.85$ & -- & $ 2.41$ & -- & $ 1.83$ \\
\ion{Fe}{ I} & 4202.029 & 1.49 & $-0.708$ & -- & -- & -- & -- & 120.2 $\pm$  11.6 & $ 4.70$ & -- & -- \\
\ion{Fe}{ I} & 4337.045 & 1.56 & $-1.695$ & -- & -- & -- & -- & -- & -- & 96.8 $\pm$   7.5 & $ 4.74$ \\
\ion{Fe}{ I} & 4352.735 & 2.22 & $-1.287$ & -- & -- & -- & -- & -- & -- & 66.7 $\pm$  11.5 & $ 4.49$ \\
\ion{Fe}{ I} & 4430.614 & 2.22 & $-1.659$ & -- & -- & -- & -- & -- & -- & 47.0 $\pm$  13.3 & $ 4.44$ \\
\ion{Fe}{ I} & 4442.339 & 2.2 & $-1.255$ & -- & -- & -- & -- & -- & -- & 84.2 $\pm$  13.1 & $ 4.76$ \\
\ion{Fe}{ I} & 4459.117 & 2.18 & $-1.279$ & -- & -- & -- & -- & -- & -- & 76.8 $\pm$  15.2 & $ 4.59$ \\
\ion{Fe}{ I} & 4602.941 & 1.49 & $-2.209$ & -- & -- & -- & -- & -- & -- & 72.2 $\pm$   9.7 & $ 4.52$ \\
\ion{Fe}{ I} & 4733.591 & 1.49 & $-2.988$ & -- & -- & -- & -- & -- & -- & 40.0 $\pm$   7.7 & $ 4.67$ \\
\ion{Fe}{ I} & 4871.318 & 2.87 & $-0.363$ & -- & -- & -- & -- & 75.0 $\pm$  13.2 & $ 4.69$ & 78.6 $\pm$  11.9 & $ 4.50$ \\
\ion{Fe}{ I} & 4872.138 & 2.88 & $-0.567$ & -- & -- & -- & -- & -- & -- & 55.3 $\pm$  11.8 & $ 4.28$ \\
\ion{Fe}{ I} & 4890.755 & 2.88 & $-0.394$ & -- & -- & 93.2 $\pm$   7.1 & $ 4.83$ & 69.3 $\pm$   9.6 & $ 4.61$ & -- & -- \\
\ion{Fe}{ I} & 4891.492 & 2.85 & $-0.112$ & -- & -- & 110.0 $\pm$   7.4 & $ 4.85$ & 75.7 $\pm$   8.9 & $ 4.43$ & 94.4 $\pm$  18.8 & $ 4.56$ \\
\ion{Fe}{ I} & 4903.31  & 2.88 & $-0.926$ & -- & -- & 68.3 $\pm$   7.4 & $ 4.87$ & -- & -- & 51.2 $\pm$   8.2 & $ 4.56$ \\
\ion{Fe}{ I} & 4918.994 & 2.87 & $-0.342$ & -- & -- & 114.4 $\pm$   3.6 & $ 5.17$ & 70.4 $\pm$  11.0 & $ 4.57$ & -- & -- \\
\ion{Fe}{ I} & 4920.502 & 2.83 & $ 0.068$ & -- & -- & -- & -- & -- & -- & 98.1 $\pm$  10.9 & $ 4.43$ \\
\ion{Fe}{ I} & 4938.814 & 2.88 & $-1.077$ & -- & -- & -- & -- & 39.0 $\pm$   6.2 & $ 4.71$ & -- & -- \\
\ion{Fe}{ I} & 5006.119 & 2.83 & $-0.638$ & -- & -- & -- & -- & 51.0 $\pm$  12.6 & $ 4.44$ & 71.3 $\pm$  14.8 & $ 4.57$ \\
\ion{Fe}{ I} & 5049.82 & 2.28 & $-1.355$ & 55.2 $\pm$   5.9 & $ 4.41$ & 96.9 $\pm$   7.6 & $ 5.10$ & -- & -- & 62.7 $\pm$  17.9 & $ 4.42$ \\
\ion{Fe}{ I} & 5110.413 & 0.0 & $-3.760$ & -- & -- & -- & -- & 81.4 $\pm$  11.1 & $ 4.73$ & -- & -- \\
\ion{Fe}{ I} & 5171.596 & 1.49 & $-1.793$ & -- & -- & -- & -- & -- & -- & 98.6 $\pm$  12.8 & $ 4.53$ \\
\ion{Fe}{ I} & 5191.455 & 3.04 & $-0.551$ & 62.5 $\pm$   8.6 & $ 4.64$ & -- & -- & -- & -- & -- & -- \\
\ion{Fe}{ I} & 5192.344 & 3.0 & $-0.421$ & 65.2 $\pm$  13.9 & $ 4.51$ & -- & -- & -- & -- & 61.3 $\pm$   9.9 & $ 4.35$ \\
\ion{Fe}{ I} & 5194.941 & 1.56 & $-2.090$ & 71.3 $\pm$  11.8 & $ 4.52$ & -- & -- & 75.4 $\pm$  13.4 & $ 4.83$ & 70.1 $\pm$  15.1 & $ 4.35$ \\
\ion{Fe}{ I} & 5202.336 & 2.18 & $-1.838$ & -- & -- & 89.2 $\pm$   5.0 & $ 5.26$ & -- & -- & 54.3 $\pm$   9.9 & $ 4.60$ \\
\ion{Fe}{ I} & 5216.274 & 1.61 & $-2.150$ & -- & -- & 80.9 $\pm$   5.4 & $ 4.68$ & -- & -- & 55.6 $\pm$   8.2 & $ 4.21$ \\
\ion{Fe}{ I} & 5217.389 & 3.21 & $-1.070$ & -- & -- & 40.2 $\pm$   7.2 & $ 4.88$ & -- & -- & -- & -- \\
\ion{Fe}{ I} & 5232.94  & 2.94 & $-0.058$ & -- & -- & 103.2 $\pm$   6.1 & $ 4.70$ & -- & -- & -- & -- \\
\ion{Fe}{ I} & 5266.555 & 3.0 & $-0.386$ & -- & -- & 86.3 $\pm$   4.2 & $ 4.76$ & 56.0 $\pm$  10.3 & $ 4.45$ & -- & -- \\
\ion{Fe}{ I} & 5281.79  & 3.04 & $-0.834$ & -- & -- & 68.8 $\pm$   6.7 & $ 4.93$ & -- & -- & 37.3 $\pm$   8.1 & $ 4.37$ \\
\ion{Fe}{ I} & 5283.621 & 3.24 & $-0.432$ & -- & -- & 71.4 $\pm$   4.6 & $ 4.83$ & -- & -- & 44.2 $\pm$   9.2 & $ 4.35$ \\
\ion{Fe}{ I} & 5302.3   & 3.28 & $-0.720$ & -- & -- & 49.6 $\pm$   4.2 & $ 4.78$ & -- & -- & -- & -- \\
\ion{Fe}{ I} & 5324.179 & 3.21 & $-0.103$ & 53.5 $\pm$   9.6 & $ 4.24$ & -- & -- & 59.3 $\pm$  11.8 & $ 4.48$ & 61.8 $\pm$   5.9 & $ 4.29$ \\
\ion{Fe}{ I} & 5332.899 & 1.56 & $-2.777$ & -- & -- & 67.5 $\pm$   4.8 & $ 4.97$ & -- & -- & 35.6 $\pm$   7.0 & $ 4.41$ \\
\ion{Fe}{ I} & 5339.929 & 3.27 & $-0.647$ & -- & -- & 54.6 $\pm$   3.9 & $ 4.77$ & -- & -- & -- & -- \\
\ion{Fe}{ I} & 5369.961 & 4.37 & $ 0.536$ & -- & -- & 43.9 $\pm$   5.2 & $ 4.76$ & -- & -- & -- & -- \\
\ion{Fe}{ I} & 5371.489 & 0.96 & $-1.645$ & 126.3 $\pm$   4.7 & $ 4.41$ & -- & -- & 114.8 $\pm$  10.5 & $ 4.54$ & -- & -- \\
\ion{Fe}{ I} & 5383.369 & 4.31 & $ 0.645$ & -- & -- & 68.4 $\pm$   6.9 & $ 5.02$ & -- & -- & -- & -- \\
\ion{Fe}{ I} & 5393.167 & 3.24 & $-0.715$ & -- & -- & 48.4 $\pm$   6.7 & $ 4.69$ & -- & -- & -- & -- \\
\ion{Fe}{ I} & 5410.91  & 4.47 & $ 0.398$ & -- & -- & 50.6 $\pm$   4.1 & $ 5.14$ & -- & -- & -- & -- \\
\ion{Fe}{ I} & 5415.199 & 4.39 & $ 0.642$ & -- & -- & 65.6 $\pm$   9.3 & $ 5.06$ & -- & -- & -- & -- \\
\ion{Fe}{ I} & 5424.068 & 4.32 & $ 0.520$ & -- & -- & 59.2 $\pm$   4.4 & $ 4.99$ & -- & -- & 38.0 $\pm$   6.5 & $ 4.61$ \\
\ion{Fe}{ I} & 5434.523 & 1.01 & $-2.122$ & 112.7 $\pm$   7.0 & $ 4.66$ & -- & -- & -- & -- & -- & -- \\
\ion{Fe}{ I} & 6065.482 & 2.61 & $-1.530$ & -- & -- & 69.0 $\pm$  12.6 & $ 5.01$ & -- & -- & 34.3 $\pm$  10.1 & $ 4.41$ \\
\ion{Fe}{ I} & 6136.615 & 2.45 & $-1.400$ & 42.7 $\pm$   7.9 & $ 4.37$ & -- & -- & -- & -- & 69.5 $\pm$   6.4 & $ 4.68$ \\
\ion{Fe}{ I} & 6137.691 & 2.59 & $-1.403$ & -- & -- & -- & -- & -- & -- & 57.1 $\pm$   8.4 & $ 4.64$ \\
\ion{Fe}{ I} & 6191.558 & 2.43 & $-1.417$ & 72.0 $\pm$   6.8 & $ 4.83$ & -- & -- & -- & -- & 62.7 $\pm$  12.3 & $ 4.55$ \\
\ion{Fe}{ I} & 6230.722 & 2.56 & $-1.281$ & 51.7 $\pm$   6.8 & $ 4.52$ & -- & -- & -- & -- & 58.2 $\pm$  10.4 & $ 4.50$ \\
\ion{Fe}{ I} & 6252.555 & 2.4 & $-1.687$ & -- & -- & 76.9 $\pm$  15.0 & $ 5.02$ & -- & -- & 42.8 $\pm$  10.2 & $ 4.45$ \\
\ion{Fe}{ I} & 6393.601 & 2.43 & $-1.432$ & -- & -- & -- & -- & -- & -- & 50.1 $\pm$  11.5 & $ 4.34$ \\
\ion{Fe}{ I} & 6400.0   & 3.6 & $-0.290$ & -- & -- & -- & -- & -- & -- & 31.5 $\pm$   3.9 & $ 4.34$ \\
\ion{Fe}{ I} & 6421.35  & 2.28 & $-2.027$ & -- & -- & 63.7 $\pm$  10.7 & $ 4.96$ & -- & -- & -- & -- \\
\ion{Fe}{ I} & 6430.845 & 2.18 & $-2.006$ & -- & -- & -- & -- & -- & -- & 48.0 $\pm$  12.7 & $ 4.55$ \\
\ion{Fe}{ I} & 6494.98  & 2.4 & $-1.273$ & -- & -- & -- & -- & -- & -- & 62.3 $\pm$  11.2 & $ 4.33$ \\
\ion{Fe}{ I} & 6677.985 & 2.69 & $-1.418$ & -- & -- & -- & -- & -- & -- & 41.5 $\pm$   8.0 & $ 4.49$ \\
\ion{Fe}{II} & 4522.627 & 2.84 & $-2.030$ & -- & -- & -- & -- & -- & -- & 45.2 $\pm$  12.2 & $ 4.26$ \\
\ion{Fe}{II} & 4583.829 & 2.81 & $-1.860$ & -- & -- & -- & -- & -- & -- & 78.6 $\pm$  13.3 & $ 4.72$ \\
\ion{Fe}{II} & 4923.921 & 2.89 & $-1.320$ & ( 88.1) & $ 4.53$ & 109.4 $\pm$   6.7 & $ 5.06$ & -- & -- & 97.7 $\pm$   6.9 & $ 4.62$ \\
\ion{Fe}{II} & 5018.436 & 2.89 & $-1.220$ & (104.9) & $ 4.76$ & -- & -- & -- & -- & 108.2 $\pm$  13.6 & $ 4.72$ \\
\ion{Fe}{II} & 5197.567 & 3.23 & $-2.100$ & -- & -- & -- & -- & -- & -- & 56.6 $\pm$  17.0 & $ 4.96$ \\
\ion{Fe}{II} & 5234.623 & 3.22 & $-2.230$ & -- & -- & 52.6 $\pm$  10.1 & $ 5.17$ & -- & -- & -- & -- \\
\ion{Fe}{II} & 5275.997 & 3.2 & $-1.940$ & -- & -- & -- & -- & -- & -- & 32.2 $\pm$   8.1 & $ 4.30$ \\
\ion{Mg}{ I} & 5172.684 & 2.71 & $-0.450$ & (162.6) & $ 4.88$ & (209.5) & $ 5.36$ & (132.0) & $ 4.92$ & (151.8) & $ 4.80$ \\
\ion{Mg}{ I} & 5183.604 & 2.72 & $-0.239$ & (173.6) & $ 5.07$ & (158.6) & $ 5.16$ & (181.3) & $ 5.25$ & (174.3) & $ 5.02$ \\
\Tstrut
\ion{Mn}{ I} & 4030     & 0.0 & $-0.494$ & -- & $ 1.55$ & -- & -- & -- & $ 2.16$ & -- & $ 1.90$ \\
\ion{Na}{ I} & 5889.951 & 0.0 & $ 0.108$ & (132.4) & $ 3.80$ & (226.7) & $ 4.16$ & (224.7) & $ 4.52$ & (193.5) & $ 3.78$ \\
\ion{Na}{ I} & 5895.924 & 0.0 & $-0.194$ & -- & -- & (174.9) & $ 4.09$ & (169.7) & $ 4.35$ & (183.4) & $ 4.00$ \\
\ion{Ni}{ I} & 5476.904 & 1.83 & $-0.780$ & -- & -- & 96.9 $\pm$   7.5 & $ 3.61$ & -- & -- & 74.8 $\pm$   7.4 & $ 3.17$ \\
\ion{Sr}{II} & 4077.709 & 0.0 & $ 0.167$ & -- & -- & (240.9) & $ 0.33$ & (124.7) & $-0.53$ & (143.1) & $-1.06$ \\
\ion{Ti}{ I} & 4681.909 & 0.05 & $-1.030$ & -- & -- & ( 60.1) & $ 2.68$ & -- & -- & -- & -- \\
\ion{Ti}{ I} & 4999.503 & 0.83 & $ 0.320$ & -- & -- & ( 74.1) & $ 2.64$ & -- & -- & -- & -- \\
\ion{Ti}{ I} & 5064.653 & 0.05 & $-0.940$ & -- & -- & ( 75.5) & $ 2.57$ & -- & -- & -- & -- \\
\ion{Ti}{II} & 4468.493 & 1.13 & $-0.630$ & -- & -- & 143.9 $\pm$  14.2 & $ 3.05$ & 110.1 $\pm$  22.7 & $ 2.52$ & 122.8 $\pm$  11.3 & $ 2.45$ \\
\ion{Ti}{II} & 4563.757 & 1.22 & $-0.690$ & 96.3 $\pm$   8.0 & $ 2.08$ & 119.1 $\pm$  13.3 & $ 2.71$ & 102.3 $\pm$  19.7 & $ 2.46$ & 100.0 $\pm$   9.4 & $ 2.08$ \\
\ion{Ti}{II} & 5336.786 & 1.58 & $-1.600$ & -- & -- & 54.0 $\pm$   3.3 & $ 2.59$ & -- & -- & -- & -- \\

\end{tabular}}
\label{Tab:lines2}
\end{table*}

\twocolumn

\begin{sidewaystable*}
\caption{Derived LTE abundances for the Fornax stars observed with UVES and the Carina stars observed with X-SHOOTER, along with their associated errors (see \S~\ref{Sec:analysis}).}
\resizebox{\linewidth}{!}{
\begin{tabular}{lrrrrrrrrrrrrrrrrrrrrrrrrrrrr}
\hline
\hline
 & \ion{Fe}{I} & \ion{Fe}{II} & C** & \ion{O}{I}  & \ion{Na}{I} & \ion{Mg}{I} & \ion{Al}{I} & \ion{Si}{I} & \ion{Ca}{I} & \ion{Sc}{II} & \ion{Ti}{I} & \ion{Ti}{II} & \ion{Cr}{I} & \ion{Mn}{I} & \ion{Co}{I} & \ion{Ni}{I} & \ion{Cu}{I} & \ion{Zn}{I} & \ion{Sr}{II} & \ion{Y}{II} & \ion{Zr}{II} & \ion{Ba}{II} & \ion{La}{II} & \ion{Pr}{II} & \ion{Nd}{II} & \ion{Eu}{II} & \ion{Dy}{II} \Tstrut \\ 
\\
log$\epsilon$(X)$_\odot$ & 7.50 & 7.50 & 8.43 & 8.69 & 6.24 & 7.60 & 6.45 & 7.51 & 6.34 & 3.15 & 4.95 & 4.95 & 5.64 & 5.43 & 4.99 & 6.22 & 4.19 & 4.56 & 2.87 & 2.21 & 2.58 & 2.18 & 1.10 & 0.72 & 1.42 & 0.52 & 1.10 \Bstrut \\
\hline

\\
fnx\_06\_019 \\
No. lines* & 67 & 2 & 1 & 1 & 2 & 3 & 1 & 1 & 6 & 6 & 9 & 7 & 4 & 3 & 1 & 3 & 1 & 1 & 1 & 5 & 1 & 4 & 4 & 1 & 3 & 2 & 3 & \\
log$\epsilon$(X) & $4.58$ & $4.67$ & $5.05$ & $6.67$ & $3.54$ & $5.23$ & $3.56$ & $5.16$ & $3.82$ & $0.40$ & $1.97$ & $2.45$ & $2.39$ & $2.09$ & $2.12$ & $3.34$ & <$0.41$ & $1.93$ & $-0.67$ & $-0.86$ & $-0.12$ & $-0.43$ & $-1.28$ & $-1.14$ & $-0.89$ & $-1.60$ & $-0.74$ & \\
{[}X/H{]} & $-2.92$ & $-2.83$ & $-3.38$ & $-2.02$ & $-2.70$ & $-2.37$ & $-2.89$ & $-2.35$ & $-2.52$ & $-2.75$ & $-2.98$ & $-2.50$ & $-3.25$ & $-3.34$ & $-2.87$ & $-2.88$ & <$-3.78$ & $-2.63$ & $-3.54$ & $-3.07$ & $-2.70$ & $-2.61$ & $-2.38$ & $-1.86$ & $-2.31$ & $-2.12$ & $-1.84$ & \\
{[}X/Fe{]} & $-0.00$ & $+0.09$ & $-0.46$ & $+0.90$ & $+0.22$ & $+0.55$ & $+0.03$ & $+0.57$ & $+0.40$ & $+0.17$ & $-0.06$ & $+0.42$ & $-0.33$ & $-0.42$ & $+0.05$ & $+0.04$ & <$-0.86$ & $+0.29$ & $-0.62$ & $-0.15$ & $+0.22$ & $+0.31$ & $+0.54$ & $+1.06$ & $+0.61$ & $+0.79$ & $+1.08$ & \\
Error & $0.10$ & $0.10$ & $0.12$ & $0.12$ & $0.10$ & $0.11$ & $0.12$ & $0.12$ & $0.10$ & $0.10$ & $0.09$ & $0.11$ & $0.09$ & $0.10$ & $0.16$ & $0.08$ & $0.12$ & $0.12$ & $0.12$ & $0.10$ & $0.12$ & $0.10$ & $0.10$ & $0.16$ & $0.10$ & $0.10$ & $0.10$ & \\
\hline

\\
fnx0579x--1 \\
No. lines* & 76 & 4 & 1 & 1 & 2 & 4 & $-$ & 1 & 11 & 5 & 10 & 8 & 5 & 3 & 1 & 4 & 1 & $-$ & 1 & 2 & $-$ & 4 & $-$ & $-$ & $-$ & $-$ & $-$ & \\
log$\epsilon$(X) & $4.77$ & $4.97$ & $5.46$ & $6.99$ & $3.50$ & $5.24$ & $-$ & $4.76$ & $3.91$ & $0.64$ & $2.01$ & $2.53$ & $2.62$ & $2.24$ & $2.07$ & $3.39$ & $0.82$ & $-$ & $-0.52$ & $-1.03$ & $-$ & $-1.21$ & $-$ & $-$ & $-$ & $-$ & $-$ & \\
{[}X/H{]} & $-2.73$ & $-2.53$ & $-2.97$ & $-1.70$ & $-2.74$ & $-2.36$ & $-$ & $-2.75$ & $-2.43$ & $-2.51$ & $-2.94$ & $-2.42$ & $-3.02$ & $-3.19$ & $-2.92$ & $-2.83$ & $-3.37$ & $-$ & $-3.39$ & $-3.24$ & $-$ & $-3.39$ & $-$ & $-$ & $-$ & $-$ & $-$ & \\
{[}X/Fe{]} & $+0.00$ & $+0.20$ & $-0.24$ & $+1.03$ & $-0.00$ & $+0.37$ & $-$ & $-0.02$ & $+0.30$ & $+0.22$ & $-0.21$ & $+0.31$ & $-0.29$ & $-0.46$ & $-0.19$ & $-0.10$ & $-0.64$ & $-$ & $-0.66$ & $-0.52$ & $-$ & $-0.66$ & $-$ & $-$ & $-$ & $-$ & $-$ & \\
Error & $0.10$ & $0.13$ & $0.12$ & $0.12$ & $0.10$ & $0.10$ & $-$ & $0.12$ & $0.10$ & $0.10$ & $0.13$ & $0.13$ & $0.10$ & $0.10$ & $0.12$ & $0.10$ & $0.12$ & $-$ & $0.12$ & $0.15$ & $-$ & $0.10$ & $-$ & $-$ & $-$ & $-$ & $-$ & \\
\hline

\\
car1\_t174 \\
No. lines* & 10 & 2 & 1 & $-$ & 1 & 2 & $-$ & $-$ & 1 & $-$ & $-$ & 1 & 1 & 1 & $-$ & $-$ & $-$ & $-$ & $-$ & $-$ & $-$ & 2 & $-$ & $-$ & $-$ & $-$ & $-$ & \\
log$\epsilon$(X) & $4.51$ & $4.64$ & $4.59$ & $-$ & $3.80$ & $4.98$ & $-$ & $-$ & $3.64$ & $-$ & $-$ & $2.08$ & $2.07$ & $1.55$ & $-$ & $-$ & $-$ & $-$ & $-$ & $-$ & $-$ & $-1.99$ & $-$ & $-$ & $-$ & $-$ & $-$ & \\
{[}X/H{]} & $-2.99$ & $-2.86$ & $-3.84$ & $-$ & $-2.44$ & $-2.63$ & $-$ & $-$ & $-2.70$ & $-$ & $-$ & $-2.87$ & $-3.57$ & $-3.88$ & $-$ & $-$ & $-$ & $-$ & $-$ & $-$ & $-$ & $-4.17$ & $-$ & $-$ & $-$ & $-$ & $-$ & \\
{[}X/Fe{]} & $-0.00$ & $+0.14$ & $-0.85$ & $-$ & $+0.55$ & $+0.37$ & $-$ & $-$ & $+0.29$ & $-$ & $-$ & $+0.12$ & $-0.58$ & $-0.89$ & $-$ & $-$ & $-$ & $-$ & $-$ & $-$ & $-$ & $-1.18$ & $-$ & $-$ & $-$ & $-$ & $-$ & \\
Error & $0.13$ & $0.13$ & $0.17$ & $-$ & $0.17$ & $0.13$ & $-$ & $-$ & $0.17$ & $-$ & $-$ & $0.17$ & $0.17$ & $0.17$ & $-$ & $-$ & $-$ & $-$ & $-$ & $-$ & $-$ & $0.26$ & $-$ & $-$ & $-$ & $-$ & $-$ & \Bstrut \\
\hline

\\
car1\_t194 \\
No. lines* & 24 & 2 & 1 & $-$ & 2 & 2 & $-$ & $-$ & 2 & $-$ & 3 & 3 & 1 & $-$ & $-$ & 1 & $-$ & $-$ & 1 & $-$ & $-$ & 2 & $-$ & $-$ & $-$ & $-$ & $-$ & \\
log$\epsilon$(X) & $4.92$ & $5.10$ & $4.90$ & $-$ & $4.12$ & $5.26$ & $-$ & $-$ & $4.10$ & $-$ & $2.63$ & $2.62$ & $2.85$ & $-$ & $-$ & $3.61$ & $-$ & $-$ & $0.33$ & $-$ & $-$ & $0.03$ & $-$ & $-$ & $-$ & $-$ & $-$ & \\
{[}X/H{]} & $-2.58$ & $-2.40$ & $-3.53$ & $-$ & $-2.12$ & $-2.34$ & $-$ & $-$ & $-2.24$ & $-$ & $-2.32$ & $-2.33$ & $-2.79$ & $-$ & $-$ & $-2.61$ & $-$ & $-$ & $-2.54$ & $-$ & $-$ & $-2.15$ & $-$ & $-$ & $-$ & $-$ & $-$ & \\
{[}X/Fe{]} & $-0.00$ & $+0.18$ & $-0.95$ & $-$ & $+0.47$ & $+0.24$ & $-$ & $-$ & $+0.34$ & $-$ & $+0.26$ & $+0.25$ & $-0.21$ & $-$ & $-$ & $-0.03$ & $-$ & $-$ & $+0.04$ & $-$ & $-$ & $+0.43$ & $-$ & $-$ & $-$ & $-$ & $-$ & \\
Error & $0.10$ & $0.16$ & $0.16$ & $-$ & $0.11$ & $0.11$ & $-$ & $-$ & $0.27$ & $-$ & $0.10$ & $0.18$ & $0.16$ & $-$ & $-$ & $0.16$ & $-$ & $-$ & $0.16$ & $-$ & $-$ & $0.11$ & $-$ & $-$ & $-$ & $-$ & $-$ & \Bstrut \\
\hline

\\
car1\_t200 \\
No. lines* & 12 & $-$ & 1 & $-$ & 2 & 2 & $-$ & $-$ & $-$ & $-$ & $-$ & 2 & 1 & 1 & $-$ & $-$ & $-$ & $-$ & 1 & $-$ & $-$ & 2 & $-$ & $-$ & $-$ & $-$ & $-$ & \\
log$\epsilon$(X) & $4.60$ & $-$ & $5.06$ & $-$ & $4.43$ & $5.08$ & $-$ & $-$ & $-$ & $-$ & $-$ & $2.49$ & $2.41$ & $2.16$ & $-$ & $-$ & $-$ & $-$ & $-0.53$ & $-$ & $-$ & $-0.60$ & $-$ & $-$ & $-$ & $-$ & $-$ & \\
{[}X/H{]} & $-2.90$ & $-$ & $-3.37$ & $-$ & $-1.80$ & $-2.52$ & $-$ & $-$ & $-$ & $-$ & $-$ & $-2.46$ & $-3.23$ & $-3.27$ & $-$ & $-$ & $-$ & $-$ & $-3.40$ & $-$ & $-$ & $-2.78$ & $-$ & $-$ & $-$ & $-$ & $-$ & \\
{[}X/Fe{]} & $+0.00$ & $-$ & $-0.47$ & $-$ & $+1.10$ & $+0.38$ & $-$ & $-$ & $-$ & $-$ & $-$ & $+0.44$ & $-0.33$ & $-0.37$ & $-$ & $-$ & $-$ & $-$ & $-0.50$ & $-$ & $-$ & $+0.12$ & $-$ & $-$ & $-$ & $-$ & $-$ & \\
Error & $0.21$ & $-$ & $0.21$ & $-$ & $0.21$ & $0.21$ & $-$ & $-$ & $-$ & $-$ & $-$ & $0.49$ & $0.21$ & $0.21$ & $-$ & $-$ & $-$ & $-$ & $0.21$ & $-$ & $-$ & $0.27$ & $-$ & $-$ & $-$ & $-$ & $-$ & \Bstrut \\
\hline

\\
LG04c\_0008 \\
No. lines* & 35 & 6 & 1 & $-$ & 2 & 2 & $-$ & $-$ & 1 & $-$ & $-$ & 2 & 1 & 1 & 2 & 1 & $-$ & $-$ & 1 & $-$ & $-$ & 2 & $-$ & $-$ & $-$ & $-$ & $-$ & \\
log$\epsilon$(X) & $4.47$ & $4.53$ & $4.22$ & $-$ & $3.89$ & $4.91$ & $-$ & $-$ & $3.55$ & $-$ & $-$ & $2.25$ & $1.83$ & $1.90$ & $1.78$ & $3.17$ & $-$ & $-$ & $-1.06$ & $-$ & $-$ & $-2.02$ & $-$ & $-$ & $-$ & $-$ & $-$ & \\
{[}X/H{]} & $-3.03$ & $-2.97$ & $-4.21$ & $-$ & $-2.35$ & $-2.69$ & $-$ & $-$ & $-2.79$ & $-$ & $-$ & $-2.70$ & $-3.81$ & $-3.53$ & $-3.21$ & $-3.05$ & $-$ & $-$ & $-3.93$ & $-$ & $-$ & $-4.20$ & $-$ & $-$ & $-$ & $-$ & $-$ & \\
{[}X/Fe{]} & $-0.00$ & $+0.06$ & $-1.18$ & $-$ & $+0.68$ & $+0.34$ & $-$ & $-$ & $+0.24$ & $-$ & $-$ & $+0.33$ & $-0.78$ & $-0.50$ & $-0.18$ & $-0.02$ & $-$ & $-$ & $-0.90$ & $-$ & $-$ & $-1.17$ & $-$ & $-$ & $-$ & $-$ & $-$ & \\
Error & $0.17$ & $0.21$ & $0.17$ & $-$ & $0.17$ & $0.17$ & $-$ & $-$ & $0.17$ & $-$ & $-$ & $0.23$ & $0.17$ & $0.17$ & $0.17$ & $0.17$ & $-$ & $-$ & $0.17$ & $-$ & $-$ & $0.22$ & $-$ & $-$ & $-$ & $-$ & $-$ & \Bstrut \\
\hline

\end{tabular}}
\label{Tab:abundances}
\tablefoot{* Number of lines kept after a careful selection of the best fit or synthesized lines. ** Measured from the CH molecular G-band.}
\end{sidewaystable*}

\onecolumn
\section{Measured lines of r-process elements}

\begin{figure*}[ht]
 \centering
  \includegraphics[height=0.2\textheight]{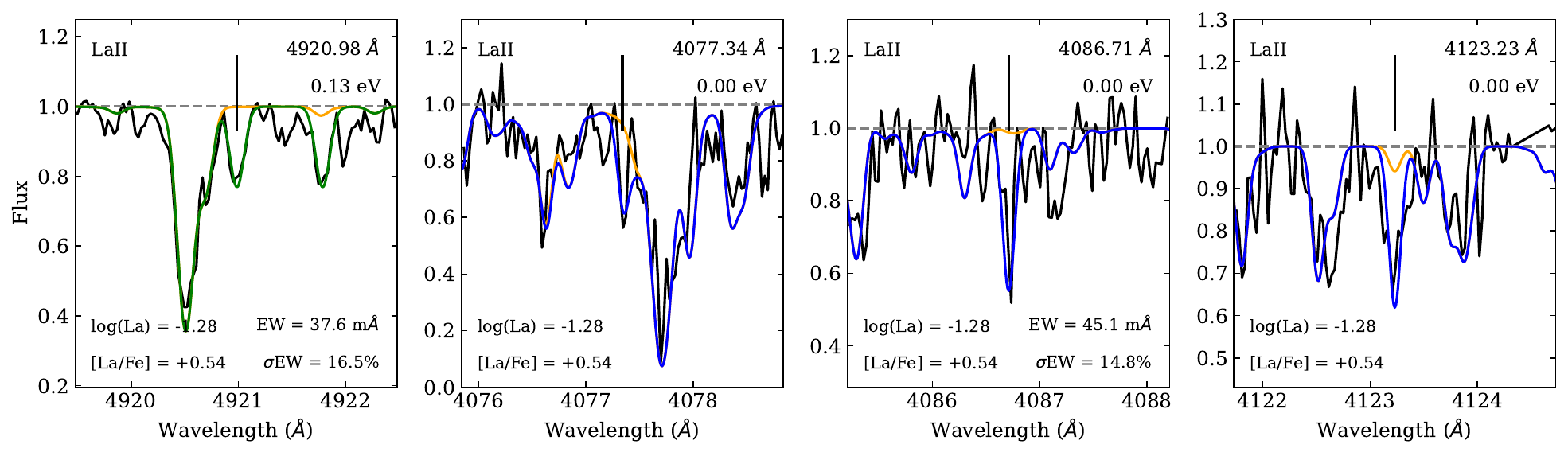}\\
  \includegraphics[height=0.2\textheight]{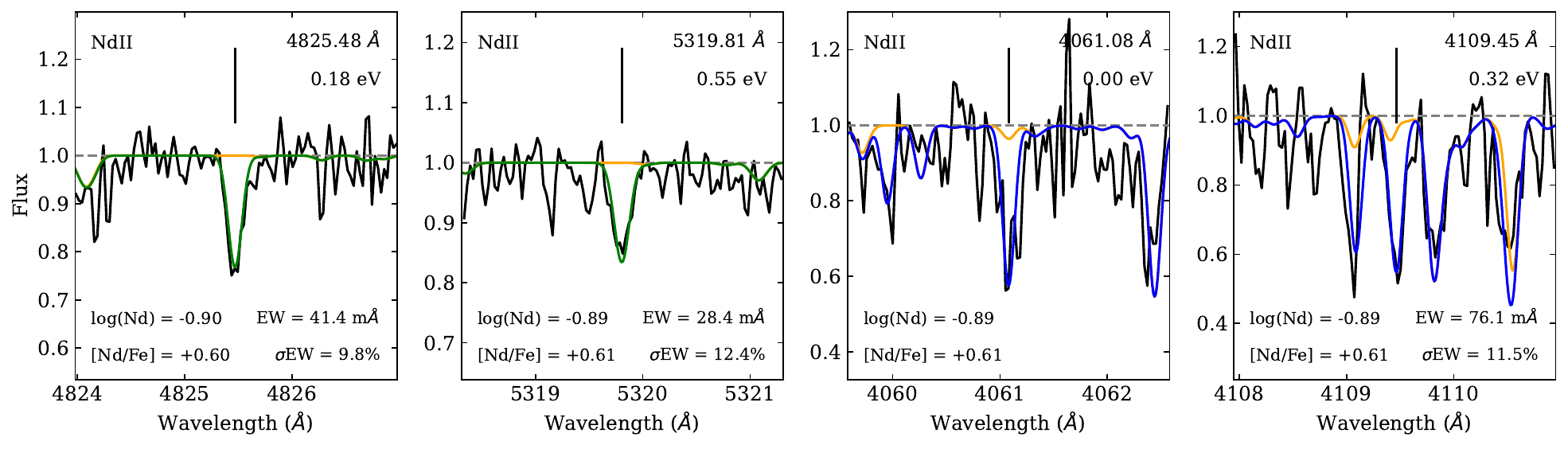}\\
  \includegraphics[height=0.2\textheight]{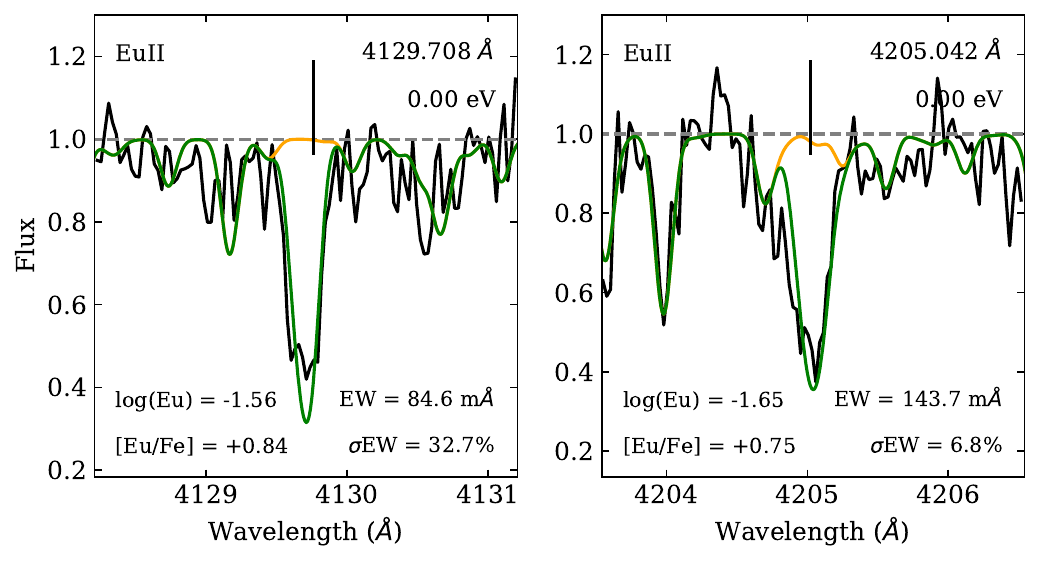}\\
  \includegraphics[height=0.2\textheight]{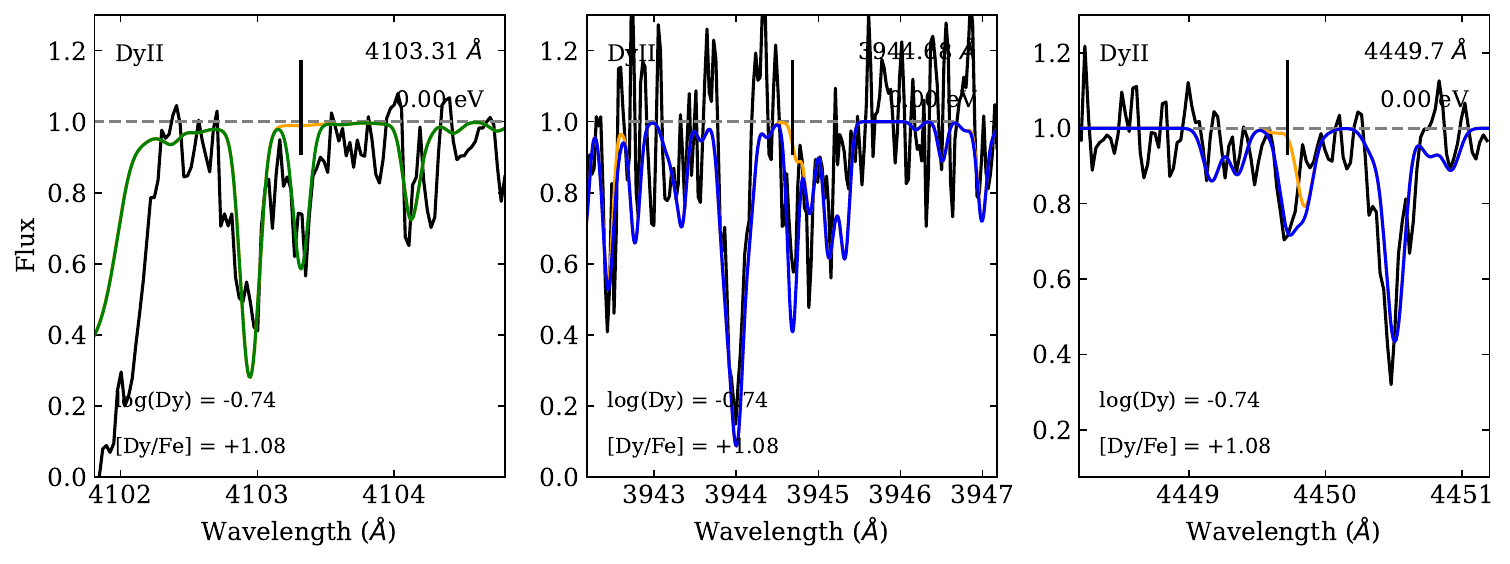}
  \caption{Individual lines of heavy (Z>56) neutron-capture elements measured in the r-process rich star, fnx\_06\_019. From top to bottom: \ion{La}{II}, \ion{Nd}{II}, \ion{Eu}{II}, and \ion{Dy}{II}. The best fit synthetic spectra used for   abundance determination are shown in green, while the synthetic spectra  computed only to check agreement with other detected lines of lesser quality are in blue. The synthetic spectra  computed without the element of interest, allowing the identification of blends, are in orange.}
  \label{Fig:individual_lines}
\end{figure*}

\twocolumn
        
\end{appendix}

\end{document}